\documentclass[nofootinbib,aps,a4paper,letterpaper,superscriptaddress,
onecolumn,times,eqsecnum]{revtex4-2}
\usepackage{amsmath}
\usepackage{amsfonts}
\usepackage{booktabs} 
\usepackage{multirow}
\usepackage{siunitx} 
\usepackage{tabularx}
\usepackage{cancel}
\usepackage{dblfloatfix}
\usepackage{float}
\usepackage[figtopcap]{subfigure}
\usepackage{array,multirow,graphicx}
\graphicspath{{./}{Figs/}}
\usepackage{adjustbox}
\usepackage{graphicx}
\usepackage{color}
\usepackage{braket}
\usepackage{dcolumn}
\usepackage{bm,url}
\usepackage[linktocpage]{hyperref}
\usepackage{subfigure}
\usepackage{amsfonts}
\usepackage[usenames,dvipsnames,svgnames]{xcolor}  
\usepackage{hyperref}   
\definecolor{oxfordblue}{rgb}{0.0, 0.13, 0.28}
\definecolor{burgundy}{rgb}{0.5, 0.0, 0.13}
\definecolor{darkolivegreen}{rgb}{0.33, 0.42, 0.18}
\definecolor{darkblue}{rgb}{0,0,0.5}
\definecolor{richcarmine}{rgb}{0.84, 0.0, 0.25}
\definecolor{darkblue}{rgb}{0,0,0.5}
\definecolor{bluer}{rgb}{0.00,0.50,0.75}{}
\hypersetup{colorlinks=true, citecolor=red, linkcolor=blue,
 urlcolor = magenta, filecolor=magenta}
\usepackage{orcidlink}

 \newcommand\be{\begin{equation}}
  \newcommand\ee{\end{equation}}
 \newcommand\bea{\begin{eqnarray}}
  \newcommand\eea{\end{eqnarray}}
 \newcommand\bseq{\begin{subequations}} 
  \newcommand\eseq{\end{subequations}}
 \newcommand\bcas{\begin{cases}}
  \newcommand\ecas{\end{cases}}

\newcommand{\beq}{\begin{equation}}
\newcommand{\eeq}{\end{equation}}

\newcommand{\comm}[1]{}

\begin{document}

\title{Rotating traversable wormholes and particle dynamics in $f(R,T)$ gravity}

\author{G.G.L. Nashed\,
\orcidlink{0000-0001-5544-1119}
}
\email{nashed@bue.edu.eg}
\affiliation{Centre for Theoretical Physics, The British University in Egypt, 
P.O. Box 43, El Sherouk City, Cairo 11837, Egypt}
\author{Waleed El 
Hanafy\,
\orcidlink{0000-0002-0097-6412}
}
\email{waleed.elhanafy@bue.edu.eg}
\affiliation{Centre for Theoretical Physics, The British University in Egypt, 
P.O. Box 43, El Sherouk City, Cairo 11837, Egypt}
\author{Amare 
Abebe\,
\orcidlink{0000-0001-5475-2919}
}
\email{Amare.Abebe@nithecs.ac.za}
\affiliation{Centre for Space Research, North-West University, Potchefstroom 
2520, South Africa}
\affiliation{National Institute for Theoretical and Computational Sciences 
(NITheCS), South Africa}

\author{Kazuharu 
Bamba\,
\orcidlink{0000-0001-9720-8817}
}
\email{bamba@sss.fukushima-u.ac.jp}
\affiliation{Faculty of Symbiotic Systems Science, Fukushima University, 
Fukushima 960-1296, Japan}

\author{Emmanuel N. Saridakis\,
\orcidlink{0000-0003-1500-0874 }
}
\email{msaridak@noa.gr}
\affiliation{National Observatory of Athens, Lofos Nymfon, 11852 Athens, Greece}
\affiliation{
Departamento de Matem\'{a}ticas, 
Universidad Cat\'{o}lica del Norte, Avda. Angamos 0610, Casilla 1280 
Antofagasta, Chile} 
\affiliation{CAS Key Laboratory for Researches in Galaxies and 
Cosmology, Department of Astronomy, University of Science and Technology of 
China, Hefei, Anhui 230026, P.R. China}

\begin{abstract}

Traversable wormholes are among the most interesting solutions of gravitational 
theories, but within General Relativity they generally require exotic matter 
violating the null energy condition. Modified gravity theories with 
matter-geometry coupling provide a promising framework in which wormhole 
geometries may instead be supported by effective gravitational contributions. 
Motivated by this possibility, we investigate rotating traversable wormholes in 
$f(R,T)$ gravity, where $R$ is the scalar curvature and $T$ is the trace of the 
energy-momentum tensor, within the slow-rotation approximation.
We construct stationary and axisymmetric wormhole solutions supported by an 
anisotropic fluid and show that the obtained geometries are regular, 
asymptotically flat, horizonless, and satisfy the flare-out condition at the 
throat. A central result is that the matter sector satisfies both the null and 
strong energy conditions, indicating that traversable rotating wormholes can be 
supported without exotic matter.
We further analyze particle motion, frame dragging, and non-geodesic effects 
arising from matter-geometry coupling, together with shadow deformation and 
gravitational lensing signatures induced by rotation. A preliminary stability 
analysis based on sound-speed conditions indicates the physical viability of 
the 
solutions. These results demonstrate that rotating wormholes in $f(R,T)$ 
gravity constitute physically consistent compact configurations with 
potentially 
observable astrophysical signatures.

\end{abstract}

 \maketitle

\section{Introduction}

General Relativity (GR) has been very successful in describing 
gravitational phenomena across a wide range of astrophysical and cosmological 
scales. Nevertheless, several fundamental issues, including the cosmological 
constant problem and the nature of 
dark energy \cite{Weinberg:1988cp}, possible observational tensions 
\cite{Abdalla:2022yfr,CosmoVerseNetwork:2025alb}, and the 
search for a consistent quantum description of gravity  \cite{Addazi:2021xuf}, 
have motivated the 
investigation of modified theories of gravity beyond Einstein's framework 
\cite{CANTATA:2021asi,Clifton:2011jh,Nojiri:2010wj,Capozziello:2011et}. 
Hence, over the last decades, 
numerous modified-gravity scenarios have been proposed, including $f(R)$  
\cite{Starobinsky:1980te, Capozziello:2002rd, DeFelice:2010aj} 
gravity,   Gauss-Bonnet and 
Lovelock gravities \cite{Nojiri:2005jg, 
DeFelice:2008wz,Asimakis:2022mbe,Lovelock:1971yv}, entropic gravities 
 \cite{Verlinde:2010hp,Carroll:2016lku,Luciano:2026ufu,Leizerovich:2026pfy},
teleparallel and $f(T)$ gravity \cite{Cai:2015emx, 
  Linder:2010py, Chen:2010va},  non-metric and $f(Q)$ gravities  
\cite{BeltranJimenez:2017tkd, Heisenberg:2023lru,Anagnostopoulos:2021ydo}, etc. 
  
 Among the various modified gravity classes, $f(R,T)$ gravity 
has emerged as a 
particularly interesting extension,  which incorporates
 non-minimal matter couplings  
 \cite{Harko:2011kv}. In particular, in this theory 
the gravitational 
Lagrangian depends not only on the Ricci scalar $R$ but also on the trace $T$ 
of the energy-momentum tensor. This explicit dependence on 
$T$ introduces a non-minimal coupling between matter and geometry, leading to 
important modifications of the gravitational dynamics and, in general, to the 
non-conservation of the energy-momentum tensor 
\cite{Harko:2011kv,Shabani:2013djy,Alvarenga:2013syu}.
Consequently, additional 
forces may appear and the motion of massive particles may deviate from geodesic 
motion. Due to its rich phenomenology, $f(R,T)$ gravity has extensively been 
 studied within the cosmological framework \cite{Jamil:2011ptc, 
  Sharif:2012zzd, Shabani:2014xvi,  Harko:2014aja,
 Myrzakulov:2012qp, Zaregonbadi:2016xna, Moraes:2015kka, Moraes:2015dee,
Alvarenga:2012bt, Chakraborty:2012kj, Bhatti:2020rzr, Barrientos:2018cnx, 
Shabani:2017kis,   Maurya:2021aio, Gamonal:2020itt, 
Baffou:2015dna,  Hansraj:2018jzb, Zubair:2020poe, 
Sahoo:2014woa, Houndjo:2012hj, Singh:2023gxd, Kumar:2021vqa, Pinto:2022tlu, 
  Pradhan:2023qzf, Hazarika:2025lln, 
Siggia:2025pvc, Farias:2026eyt}.

On the other hand, compact solutions of the gravitational field equations, such 
as black holes and 
wormholes, constitute particularly important probes of gravitational theories 
in 
the strong-field regime. Black holes play a central role in modern 
astrophysics, 
gravitational-wave physics, and high-energy phenomena, while the direct imaging 
of black-hole shadows by the Event Horizon Telescope has opened a new era for 
testing gravity observationally \cite{Barack:2018yly}. Similarly,  wormholes 
represent 
nontrivial spacetime configurations connecting distinct regions of spacetime 
and 
have attracted considerable attention both from theoretical and observational 
perspectives. Originally introduced through the Einstein-Rosen bridge 
\cite{Einstein:1935tc}, traversable wormholes were later formulated by Morris 
and Thorne \cite{Morris:1988cz} as geometries allowing in principle the passage 
of matter and signals between two asymptotically flat regions. However, within 
GR, traversable wormholes generically require exotic matter violating the null 
energy condition (NEC), which constitutes one of the main obstacles to their 
physical viability \cite{Visser1995,Lobo:2005uf}. This difficulty has motivated 
extensive 
investigation of wormhole solutions in modified gravity, where effective 
geometrical contributions may support the wormhole structure without requiring 
pathological matter sources. 

In addition, rotation constitutes a 
fundamental feature of realistic astrophysical compact objects. Rotating 
wormholes exhibit frame-dragging effects, modified orbital dynamics, and richer 
observational signatures compared to static configurations 
\cite{Teo:1998dp,Kashargin:2007mm}. In particular, rotation can significantly 
influence the 
motion of massive particles and photons, affecting effective potentials, stable 
orbits, gravitational lensing, and shadow formation. Nevertheless, exact 
rotating wormhole solutions are generally difficult to obtain analytically, and 
thus 
the slow-rotation approximation provides a physically well-motivated framework 
for investigating the leading rotational effects while maintaining analytic 
control over the geometry.

Based on these considerations, in the present work we investigate rotating 
traversable wormhole solutions within the framework of $f(R,T)$ gravity.
Unlike most previous   wormhole studies in $f(R,T)$ gravity, 
which   
focused on static configurations \cite{Moraes:2016jyi, Zubair:2016cde, 
Moraes:2017rrv, 
Deb:2017rhc, 
Moraes:2017zgm, Singh:2018xjv, Sharif:2018khl, Das:2017rhi, Yousaf:2018jkb, 
Banerjee:2020uyi, Tangphati:2022mur, Azmat:2022bbc}, the present analysis 
incorporates rotational 
effects and investigates their impact on particle dynamics, frame dragging, 
shadow formation, and gravitational lensing within a unified framework. 
Furthermore, the non-geodesic corrections induced by the matter-geometry 
coupling are explicitly analyzed through the extra-force contribution, 
providing 
additional insight into the dynamical and observational properties of rotating 
wormholes in modified gravity.
  Our results demonstrate that the matter-geometry coupling in $f(R,T)$ gravity 
can naturally support physically consistent rotating wormhole geometries, while 
simultaneously inducing characteristic dynamical and observational effects that 
distinguish them from their GR counterparts.

The remainder of this work is organized as follows. In 
Sec.~\ref{sec:framework}, we present the framework of $f(R,T)$ gravity and 
derive 
the field equations for rotating wormhole geometries in the slow-rotation 
regime. In Sec.~\ref{sec:solutions}, we construct explicit rotating wormhole 
solutions and analyze the corresponding matter sector and energy conditions. 
The 
geometrical properties and stability of the obtained solutions are investigated 
in Sec.~\ref{sec:geometry_stability}. In 
Sec.~\ref{sec:dynamics}, we examine particle motion, frame dragging, and the 
extra-force contribution induced by the matter-geometry coupling. The 
observational signatures associated with shadows and gravitational lensing are 
studied in Sec.~\ref{sec:observations}. Finally, the main conclusions are 
summarized in Sec.~\ref{sec:conclusions}.

\section{$f(R,T)$ gravity and rotating wormhole framework}
\label{sec:framework}

In this section, we present the theoretical framework of our analysis 
and  derive the modified gravitational equations governing slowly rotating 
traversable wormholes in $f(R,T)$ gravity.   These ingredients constitute 
the basis for the construction and physical analysis of the wormhole solutions 
investigated in the following sections.

\subsection{Field equations and matter-geometry coupling}
\label{subsec:field_equations}

The $f(R,T)$ theory of gravity constitutes an important extension of General 
Relativity. In $f(R,T)$ gravity, the gravitational Lagrangian is generalized to 
depend 
on both the Ricci scalar $R$ and the trace $T$ of the energy-momentum tensor 
\cite{Harko:2011kv}. In contrast to purely curvature-based 
modifications of gravity, the explicit dependence on $T$ introduces a direct 
matter-geometry coupling, allowing the matter sector itself to contribute 
nontrivially to the modification of the gravitational dynamics. Such couplings 
may effectively arise from quantum corrections, imperfect fluids, or exotic 
matter sources, and they lead to important deviations from the standard 
Einstein 
theory, particularly in strong-field regimes and compact configurations.

The action of the theory is given by
\begin{equation}\label{act}
S = \int \left[
\frac{c^4}{16\pi G} f(R,T)
+ \mathcal{L}_\mathrm{m}
\right]
\sqrt{-g}\, d^4x,
\end{equation}
where 
$\mathcal{L}_\mathrm{m}$ is the matter Lagrangian density,   $g$ is the 
determinant of 
the metric tensor $g_{\mu\nu}$,   
  $G$ is Newton's gravitational constant, and $c$ is the speed of light
(in the following, we adopt geometrized 
units by setting $G=c=1$, unless otherwise stated).
Furthermore, the energy-momentum tensor is defined through
\begin{equation}
T_{\mu\nu}
=
-\frac{2}{\sqrt{-g}}
\frac{\delta\!\left(\sqrt{-g}\,\mathcal{L}_\mathrm{m}\right)}
{\delta g^{\mu\nu}} .
\end{equation}

Varying the action \eqref{act} with respect to the metric tensor yields the 
modified field equations \cite{Harko:2011kv}
\begin{align}\label{fe}
f_R(R,T)R_{\mu\nu}
-\frac12 f(R,T)g_{\mu\nu}
+\left(g_{\mu\nu}\Box-\nabla_\mu\nabla_\nu\right)f_R(R,T)
=
\frac{8\pi G}{c^4}T_{\mu\nu}
-f_T(R,T)\left(T_{\mu\nu}+\Theta_{\mu\nu}\right),
\end{align}
with
$
f_R(R,T)\equiv
\frac{\partial f(R,T)}{\partial R}$, $
f_T(R,T)\equiv
\frac{\partial f(R,T)}{\partial T},
$
and where
$
\Box\equiv\nabla^\alpha\nabla_\alpha
$
denotes the d'Alembertian operator. Moreover, the tensor $\Theta_{\mu\nu}$ is 
defined as
\begin{equation}
\Theta_{\mu\nu}
\equiv
g^{\alpha\beta}
\frac{\delta T_{\alpha\beta}}
{\delta g^{\mu\nu}} .
\end{equation}
Additionally, assuming that the matter Lagrangian depends only on the metric 
components and 
not on their derivatives, one obtains
\begin{equation}
\Theta_{\mu\nu}
=
-2T_{\mu\nu}
+g_{\mu\nu}\mathcal{L}_\mathrm{m}
-2g^{\alpha\beta}
\frac{\partial^2\mathcal{L}_\mathrm{m}}
{\partial g^{\mu\nu}\partial g^{\alpha\beta}} .
\end{equation}

In the present work, we focus on the linear 
model
\begin{equation}\label{lin}
f(R,T)=R+\lambda T,
\end{equation}
where $\lambda$ is a dimensional coupling constant controlling the strength of 
the matter-geometry interaction (since $[R]=L^{-2}$ and 
$[T]=L^{-4}$, dimensional consistency requires $[\lambda]=L^2$). 
This linear form represents the simplest nontrivial realization of 
matter-geometry coupling and has been widely employed in cosmological and 
astrophysical applications. 
For the choice \eqref{lin}, one has
$
f_R=1$ and $
f_T=\lambda $, and thus the field equations reduce to
\begin{equation}\label{linear_fe}
R_{\mu\nu}
-\frac12(R+\lambda T)g_{\mu\nu}
=
8\pi T_{\mu\nu}
-\lambda\left(T_{\mu\nu}+\Theta_{\mu\nu}\right).
\end{equation}
Equation \eqref{linear_fe} explicitly displays the modification of the Einstein 
equations induced by the matter-geometry coupling. In particular, the 
additional $T$-dependent terms effectively alter the gravitational sector and 
may significantly modify the conditions required to support compact and 
nontrivial spacetime geometries.

A characteristic feature of $f(R,T)$ gravity is that the energy-momentum 
tensor 
is not covariantly conserved in general. Taking the divergence of the field 
equations leads to
\begin{equation}
\nabla^\mu T_{\mu\nu}\neq0 ,
\end{equation}
which implies that massive test particles may experience an additional 
interaction induced by the matter-geometry coupling. The corresponding 
non-geodesic effects will be investigated later in the context of particle 
motion around the rotating wormhole geometry.

Finally, in order to model the matter distribution supporting the wormhole, we 
consider an 
anisotropic fluid, which is particularly appropriate for compact configurations 
and wormhole spacetimes. Its energy-momentum tensor is written as
\begin{equation}\label{anisotropic_Tmunu}
T_{\mu\nu}
=
(\rho+p_t)u_\mu u_\nu
+p_t g_{\mu\nu}
+(p_r-p_t)\chi_\mu\chi_\nu ,
\end{equation}
where $\rho$ is the energy density, $p_r$ and $p_t$ are the radial and 
tangential pressures respectively, $u^\mu$ is the four-velocity obeying
$
u^\mu u_\mu=-1,
$
and $\chi^\mu$ is a unit spacelike vector in the radial direction satisfying
$
\chi^\mu\chi_\mu=1,
$ with 
$u^\mu\chi_\mu=0 $.
We mention that throughout the manuscript we adopt the metric signature
$
(-,+,+,+).$ Hence, the trace of the energy-momentum tensor therefore becomes
\begin{equation}\label{trace_T}
T
=
g^{\mu\nu}T_{\mu\nu}
=
-\rho+p_r+2p_t .
\end{equation}

\subsection{Rotating wormhole geometry in the slow-rotation regime}
\label{subsec:rotating_geometry}

In order to investigate rotating traversable wormholes in $f(R,T)$ gravity, we 
consider a stationary and axisymmetric spacetime. Such geometries generalize 
the 
static Morris-Thorne wormhole by incorporating the off-diagonal metric 
component responsible for frame dragging. Following the construction introduced 
by Teo \cite{Teo:1998dp}, the rotating wormhole line element can be written as
\begin{align}\label{metric}
ds^2= -N^2(r)\, dt^2
+ \frac{dr^2}{1 - \frac{b(r)}{r}}
+ r^2 \left[d\theta^2 + \sin^2\theta \left(d\phi - \omega(r)\, dt \right)^2 
\right],
\end{align}
where $N(r)$ is the redshift function, $b(r)$ is the shape function, and 
$\omega(r)$ is the angular velocity associated with the dragging of inertial 
frames.

The redshift function must remain finite and nonzero everywhere in order to 
avoid horizons and ensure traversability. The shape function determines the 
spatial geometry and satisfies the throat condition
\begin{equation}\label{throat_condition}
b(r_0)=r_0 ,
\end{equation}
where $r_0$ denotes the wormhole throat radius. In addition, the flare-out 
condition requires
\begin{equation}\label{flareout_condition}
\frac{b(r)-r b'(r)}{b^2(r)} >0 ,
\end{equation}
at or near the throat \cite{Morris:1988cz,Visser1995}. Finally, asymptotic 
flatness imposes
\begin{equation}
\lim_{r\to\infty}\omega(r)=0 .
\end{equation}

Since exact rotating wormhole solutions are generally difficult to realise
analytically, we work within the slow-rotation approximation, treating the 
angular velocity as a first-order quantity,
\begin{equation}
\omega(r)=\epsilon\,\Omega(r),
\qquad
\epsilon\ll1 ,
\end{equation}
where $\epsilon$ is a dimensionless bookkeeping parameter. Expanding the metric 
up to first order in $\epsilon$, one obtains
\begin{equation}
\sin^2\theta\left(d\phi-\omega(r)\,dt\right)^2
=
\sin^2\theta
\left[
d\phi^2
-2\omega(r)\,dt\,d\phi
\right]
+\mathcal{O}(\epsilon^2),
\end{equation}
and therefore the metric becomes
\begin{align}\label{slow_metric}
ds^2 \simeq {}&
-N^2(r)\,dt^2
+\frac{dr^2}{1-\frac{b(r)}{r}}
+r^2d\theta^2
+r^2\sin^2\theta\,d\phi^2
-2r^2\sin^2\theta\,\omega(r)\,dt\,d\phi .
\end{align}
The above approximation retains the leading frame-dragging contribution while 
keeping the zeroth-order geometry identical to the static Morris-Thorne 
wormhole. In the limit $\omega(r)\to0$, Eq.~\eqref{metric} reduces to the 
standard static traversable wormhole geometry \cite{Morris:1988cz}.
Finally, for convenience, we introduce
\begin{equation}
\Delta(r)\equiv1-\frac{b(r)}{r}.
\end{equation}
Since all metric functions depend only on the radial coordinate, substitution 
of 
the slow-rotation metric \eqref{slow_metric} into the modified field equations 
\eqref{fe} yields a coupled system of differential equations governing the 
functions $N(r)$, $b(r)$, and $\omega(r)$.

For the ansatz \eqref{slow_metric}, and for the linear model  \eqref{lin}, the 
field equations \eqref{linear_fe} reduce to
\begin{align}\label{fe2}
N\left[
\Delta N''
+\left(
\frac{2}{r}
-\frac{b'}{2r}
-\frac{3b}{2r^2}
\right)N'
\right]
+\frac{1}{2}(R+\lambda T)N^2
&=
8\pi T_{tt}
-\lambda\left(T_{tt}+\Theta_{tt}\right),
\nonumber\\
\left[
-\frac{N''}{N}
+\frac{rb'-b}{2r(r-b)}\frac{N'}{N}
+\frac{rb'-b}{r^2(r-b)}
\right]
-\frac{R+\lambda T}{2\Delta}
&=
8\pi T_{rr}
-\lambda\left(T_{rr}+\Theta_{rr}\right),
\nonumber\\
R_{t\phi}
+\frac{1}{2}(R+\lambda T)\,r^2\sin^2\theta\,\omega
&=
8\pi T_{t\phi}
-\lambda\left(T_{t\phi}+\Theta_{t\phi}\right).
\end{align} 
The above system constitutes the set of modified gravitational equations 
governing slowly rotating traversable wormholes in the linear 
$f(R,T)$ model.

\section{Rotating wormhole solutions}
\label{sec:solutions}

We now proceed to construct explicit rotating wormhole solutions within the 
framework developed in the previous section. Adopting suitable choices for the 
shape function and the rotational profile, we derive asymptotically flat and 
traversable geometries compatible with the modified field equations. We further 
analyze the corresponding matter sector and examine the role of the 
matter-geometry coupling in supporting physically consistent wormhole 
configurations.

\subsection{Shape function and rotational profile}
\label{subsec:shape_rotation}

In order to construct explicit solutions, we adopt a reconstruction approach. 
Namely, we assume physically motivated wormhole metric functions satisfying the 
throat, flare-out and asymptotic-flatness requirements, and we then determine 
the matter sector required by the modified field equations. This approach is 
common in the study of wormhole geometries, where the nontrivial topology is 
imposed through the metric functions and the corresponding matter distribution 
is reconstructed consistently.

In the following, we take the redshift function to be constant,
\begin{equation}
N(r)=1 ,
\end{equation}
which ensures the absence of horizons and simplifies the field equations 
without affecting the essential wormhole properties. For the shape function, we 
consider the power-law form
\begin{equation}\label{shape_general}
b(r)=r_0\left(\frac{r_0}{r}\right)^m ,
\end{equation}
where $r_0$ is the throat radius and $m$ is a constant parameter. This choice 
satisfies the throat condition $b(r_0)=r_0$ and gives
$
b'(r_0)=-m $. 
Hence, for $m>0$ one has $b'(r_0)<1$, and the flare-out condition is satisfied 
at the throat.

The rotational profile is determined by the $(t\phi)$ component of the modified 
field equations. Assuming a slowly rotating configuration with vanishing 
off-diagonal matter source, the corresponding equation reduces to
\begin{equation}\label{omega_eq}
(r-b)\omega''
+
\left(
\frac{4r-3b-rb'}{r}
\right)\omega'=0 .
\end{equation}
For the shape function \eqref{shape_general}, Eq.~\eqref{omega_eq} gives
\begin{equation}\label{omega_solution}
\omega(r)=C_1+C_2\int
\frac{dr}{r^4\sqrt{1-\frac{b(r)}{r}}} ,
\end{equation}
where $C_1$ and $C_2$ are integration constants. Asymptotic flatness requires 
$\omega(r)\to0$ at large $r$, which fixes the constant asymptotic part. 
Moreover, since $b(r)/r\to0$ as $r\to\infty$,  \eqref{omega_solution} 
implies
\begin{equation}
\omega(r)\sim \frac{C}{r^3},
\qquad r\to\infty.
\label{asymptbehav}
\end{equation}
Additional details concerning the asymptotic behavior of the rotational 
function 
and its relation to photon circular orbits are presented in Appendix 
\ref{app:rotation}.
Thus, the rotational function exhibits the standard frame-dragging falloff, 
analogous to the Lense-Thirring behavior of slowly rotating compact objects. 
This ensures that the rotational effects are significant near the throat, while 
standard asymptotically flat behavior is recovered far from the wormhole.

\subsection{Matter sector and energy conditions}
\label{subsec:matter_energy}

We now determine the matter distribution associated with the above rotating 
wormhole geometry. For the anisotropic fluid whose energy-momentum tensor is 
described by Eq.~\eqref{anisotropic_Tmunu}, we 
assume the linear relation
\begin{equation}\label{eos}
p_t=n p_r ,
\end{equation}
where $n$ is a constant anisotropy parameter. Hence, the trace of the 
energy-momentum tensor in Eq.~(\ref{trace_T}) is written as 
\begin{equation}
T=-\rho+(1+2n)p_r .
\end{equation}

\begin{figure}[H]
\centering
\subfigure[~Shape 
Function]{\label{fig:T1}\includegraphics[width=0.36\linewidth]{
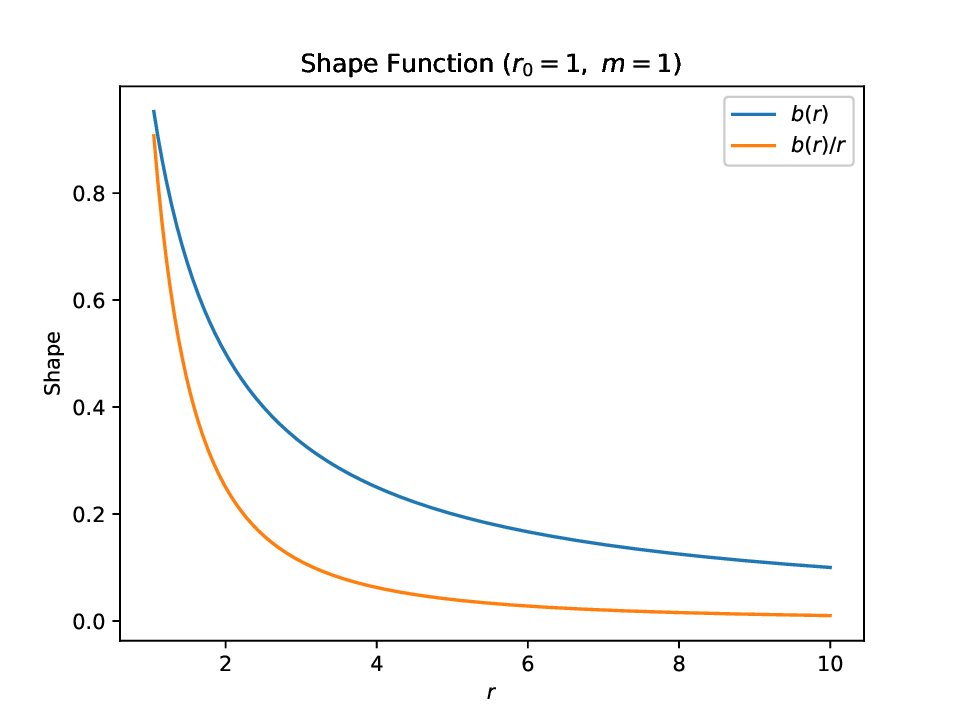}}
\subfigure[~Rotation 
Function]{\label{fig:T2}\includegraphics[width=0.36\linewidth]{
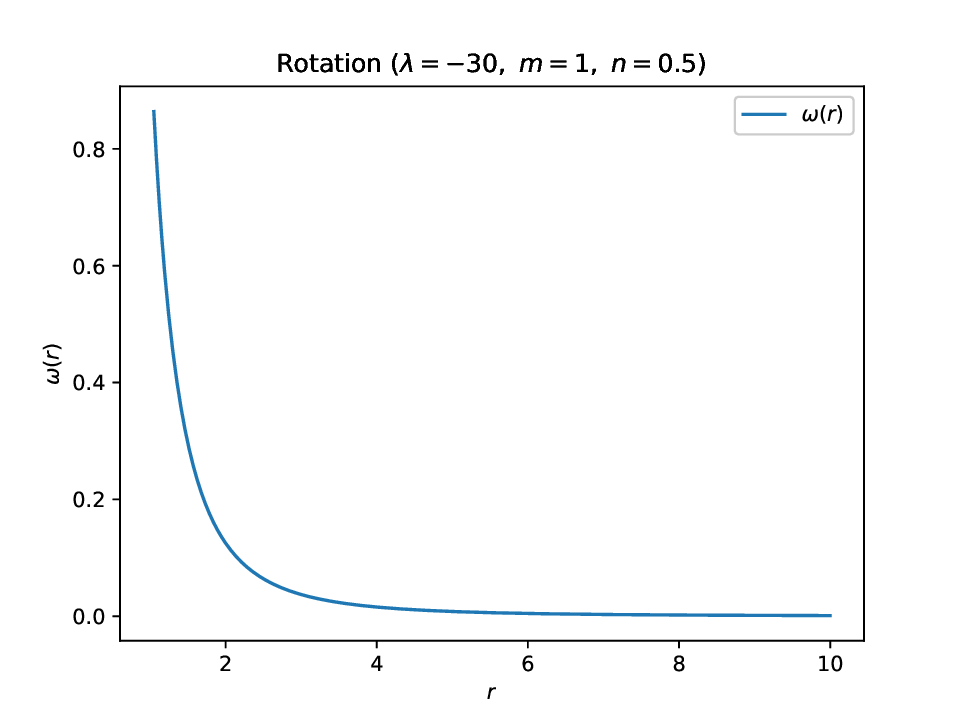}}
\subfigure[~Matter 
Components]{\label{fig:T3}\includegraphics[width=0.36\linewidth]{
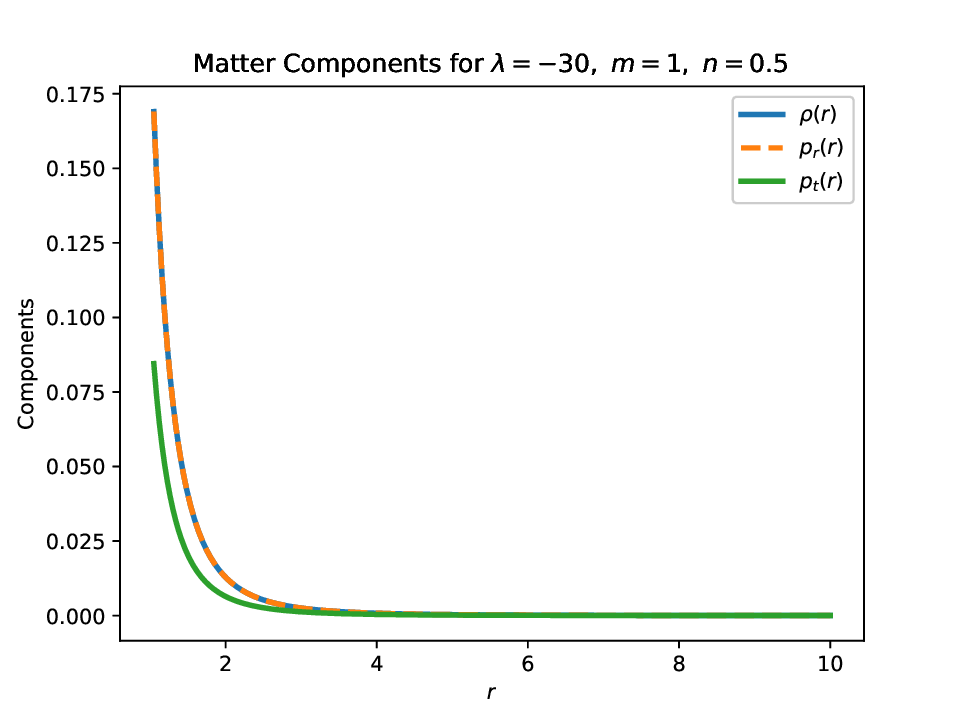}}
\subfigure[~Energy 
Conditions]{\label{fig:T4}\includegraphics[width=0.36\linewidth]{
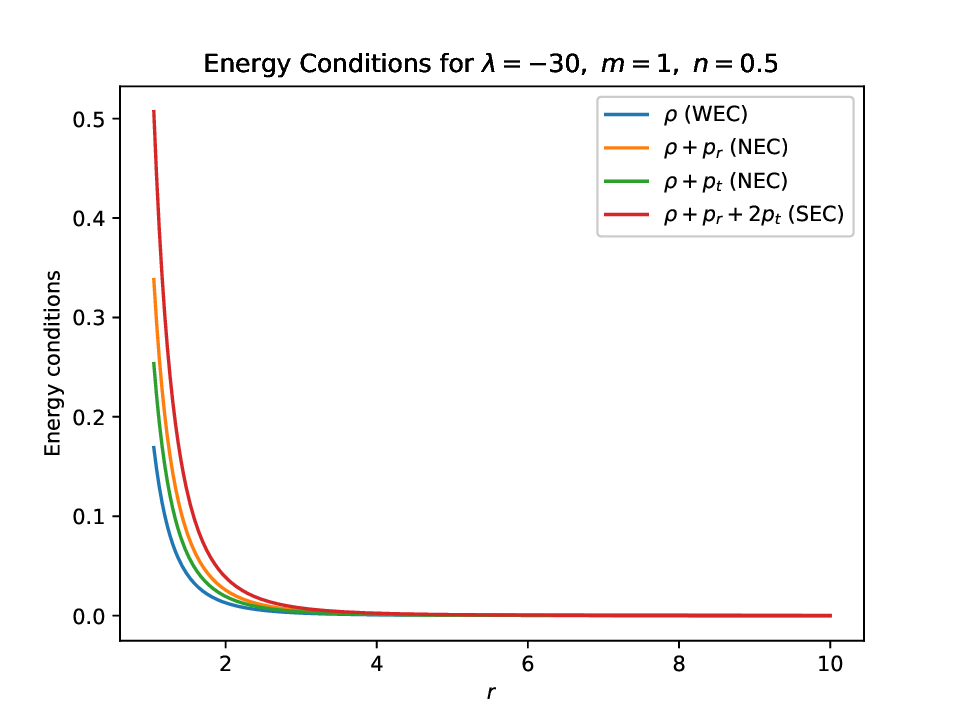}}
\caption{
{\it{
Radial behavior of the rotating wormhole solution and the corresponding matter 
sector. 
Panel (a) displays the shape function $b(r)$ together with the flare-out 
condition 
$b(r)<r$, confirming the traversable character of the geometry. 
Panel (b) shows the rotational profile $\omega(r)$, illustrating the asymptotic 
decay of frame-dragging effects. 
Panel (c) presents the matter variables $\rho$, $p_r$, and $p_t$, which remain 
positive and regular outside the throat. 
Panel (d) depicts the relevant energy-condition combinations, demonstrating 
that the null and strong energy conditions are satisfied throughout the 
spacetime.}}
}
\label{fig:1}
\end{figure}

Using the field equations for $N(r)=1$ and the shape function 
\eqref{shape_general}, one obtains
\begin{equation}\label{rho_pr_pt_general}
\rho(r)=
-\frac{m\,b(r)}{(8\pi+\lambda)r^3},
\qquad
p_r(r)=
-\frac{b(r)}{(8\pi+\lambda)r^3},
\qquad
p_t(r)=
-\frac{n\,b(r)}{(8\pi+\lambda)r^3}.
\end{equation}
Therefore, for $m>0$, $n>0$ and $b(r)>0$, the matter variables are positive 
provided that
\begin{equation}
\lambda<-8\pi .
\end{equation}
Additionally, the energy-momentum tensor then takes the compact form
\begin{equation}
T^\mu{}_\nu(r)
=
-\frac{b(r)}{(8\pi+\lambda)r^3}
\mathrm{diag}(m,1,n,n).
\end{equation}

In order to obtain physically reasonable matter distributions, for 
definiteness    we consider the 
representative configuration
\begin{equation}\label{representative_parameters}
m=1,
\qquad
n=\frac{1}{2},
\qquad
\lambda=-30 ,
\end{equation}
which satisfies the condition $\lambda<-8\pi$. The corresponding shape function 
becomes
\begin{equation}\label{shape_specific}
b(r)=r_0\left(\frac{r_0}{r}\right)=\frac{r_0^2}{r}.
\end{equation}
Under these assumptions we have
$
\rho(r)>0,
$ $
p_r(r)>0,
$ $
p_t(r)>0,
$
for all $r\ge r_0$, while all matter components decay monotonically as
\begin{equation}
\rho,\,p_r,\,p_t \sim \frac{1}{r^4},
\qquad r\to\infty ,
\end{equation}
indicating that the matter distribution is regular outside the throat and fully 
compatible with asymptotic flatness.

We now examine the standard energy conditions for the matter sector. From 
Eq.~\eqref{rho_pr_pt_general} one immediately obtains
\begin{equation}
\rho>0,
\qquad
\rho+p_r>0,
\qquad
\rho+p_t>0 ,
\end{equation}
thereby confirming that the null energy condition (NEC) is satisfied. Moreover,
\begin{equation}
\rho+p_r+2p_t>0 ,
\end{equation}
which shows that the strong energy condition (SEC) is also satisfied.

It is important to emphasize that, unlike in GR, where 
traversable wormholes generically require exotic matter violating the null 
energy condition, the present model admits wormhole solutions supported by a 
matter sector satisfying both the NEC and SEC. This behavior originates from 
the matter-geometry coupling in $f(R,T)$ gravity, which effectively modifies 
the gravitational sector and redistributes the energy contributions. As a 
result, physically acceptable matter sources can consistently support the 
wormhole geometry without the need for exotic matter in the usual GR sense.

The behavior of the shape function, rotational profile, matter components, and 
energy-condition combinations is illustrated in 
Fig.~\ref{fig:1}. As we observe, all quantities remain regular near the throat 
and decay smoothly at large radii, consistently with the traversable and 
asymptotically flat character of the obtained rotating wormhole solution.

\section{Geometrical properties and stability}
\label{sec:geometry_stability}

Having constructed the rotating wormhole solutions, we now investigate their 
geometrical properties and physical viability. In particular, we analyze the 
embedding structure of the obtained configurations in order to visualize the 
wormhole geometry and verify the traversable character of the spacetime. We 
further examine preliminary stability conditions and discuss the consistency of 
the solutions within the framework of matter-geometry-coupled gravity.

\subsection{Embedding structure and traversability}
\label{subsec:embedding}

A useful way to visualize the spatial geometry of a wormhole is through an 
embedding diagram. Considering a constant-time slice 
$(t=\mathrm{const})$ and restricting to the equatorial plane 
$(\theta=\pi/2)$, the metric reduces to
\begin{equation}
ds^2=
\frac{dr^2}{1-\frac{b(r)}{r}}
+r^2d\phi^2,
\end{equation}
which describes a two-dimensional curved surface.
This geometry can be embedded in a three-dimensional Euclidean space with 
cylindrical coordinates $(r,\phi,z)$ and line element
\begin{equation}
ds^2=dz^2+dr^2+r^2d\phi^2.
\end{equation}
Comparison of the two metrics yields the embedding condition
\begin{equation}
\frac{dz}{dr}
=
\pm\frac{1}{\sqrt{\frac{r}{b(r)}-1}}.
\label{eq:embedding}
\end{equation}
Hence, for the representative shape function
(\ref{shape_specific}), relation \eqref{eq:embedding} integrates to
\begin{equation}
z(r)=\pm\sqrt{r^2-r_0^2}.
\end{equation}
Thus, the resulting surface consists of two symmetric branches joined at the 
minimum 
radius $r=r_0$, which defines the wormhole throat. Near the throat, we have
\begin{equation}
\frac{dz}{dr}\rightarrow\infty,
\qquad
r\rightarrow r_0,
\end{equation}
indicating the characteristic vertical tangent associated with the flare-out 
behavior. Additionally, for the chosen shape function, we obtain
$
b'(r_0)=-1<1,
$
confirming that the flare-out condition is satisfied. Finally, at large 
distances we acquire
\begin{equation}
z(r)\sim r,
\qquad
r\rightarrow\infty,
\end{equation}
showing that the geometry becomes asymptotically flat on both sides of the 
throat.

The embedding diagram therefore provides a clear geometric interpretation of 
the wormhole as a smooth bridge connecting two asymptotically flat regions. 
The regular behavior of the embedded surface indicates the absence of 
geometrical singularities at the throat, while the condition $N(r)=1$ ensures 
that no event horizon is present and the configuration remains traversable in 
principle.
Unlike in GR, the flare-out condition in the present framework 
does not require violation of the NEC. As shown in the 
previous section, the matter sector satisfies both the NEC and SEC, while the 
matter-geometry coupling in $f(R,T)$ gravity effectively contributes to the 
gravitational dynamics supporting the wormhole structure.

\begin{figure}[htbp]
\centering
\subfigure[~Embedding function $z(r)=\pm\sqrt{r^2-r_0^2}$]{%
\label{fig:T5}
\includegraphics[height=4.2cm]{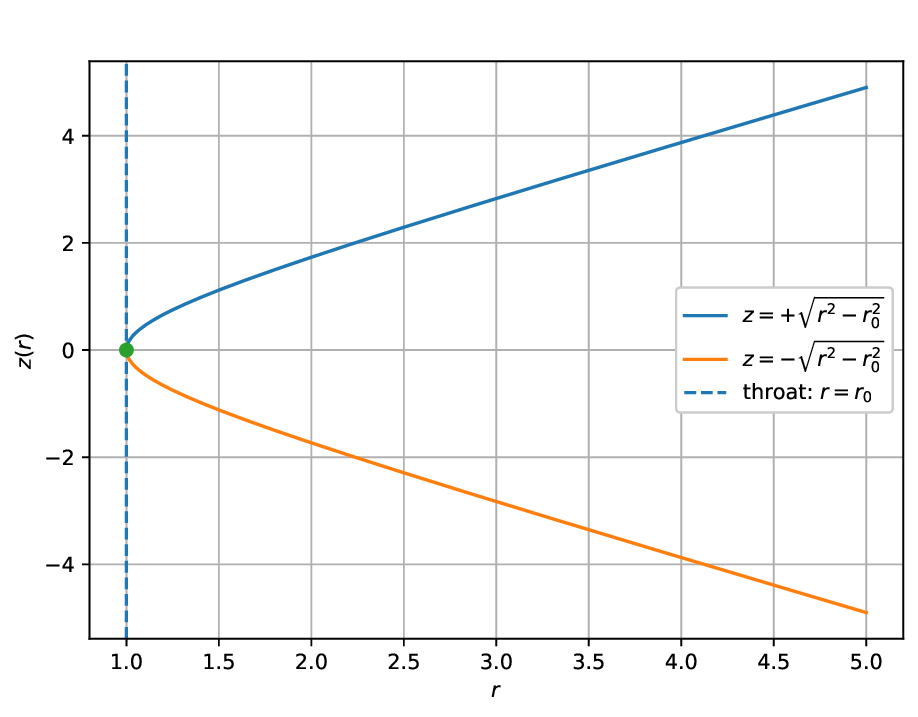}}
\hfill
\subfigure[~Derivative of the embedding function $\frac{dz}{dr}$]{%
\label{fig:T6}
\includegraphics[height=4.2cm]{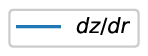}}
\hfill
\subfigure[~Three-dimensional embedding surface of the wormhole geometry]{%
\label{fig:T7}
\includegraphics[height=4.2cm]{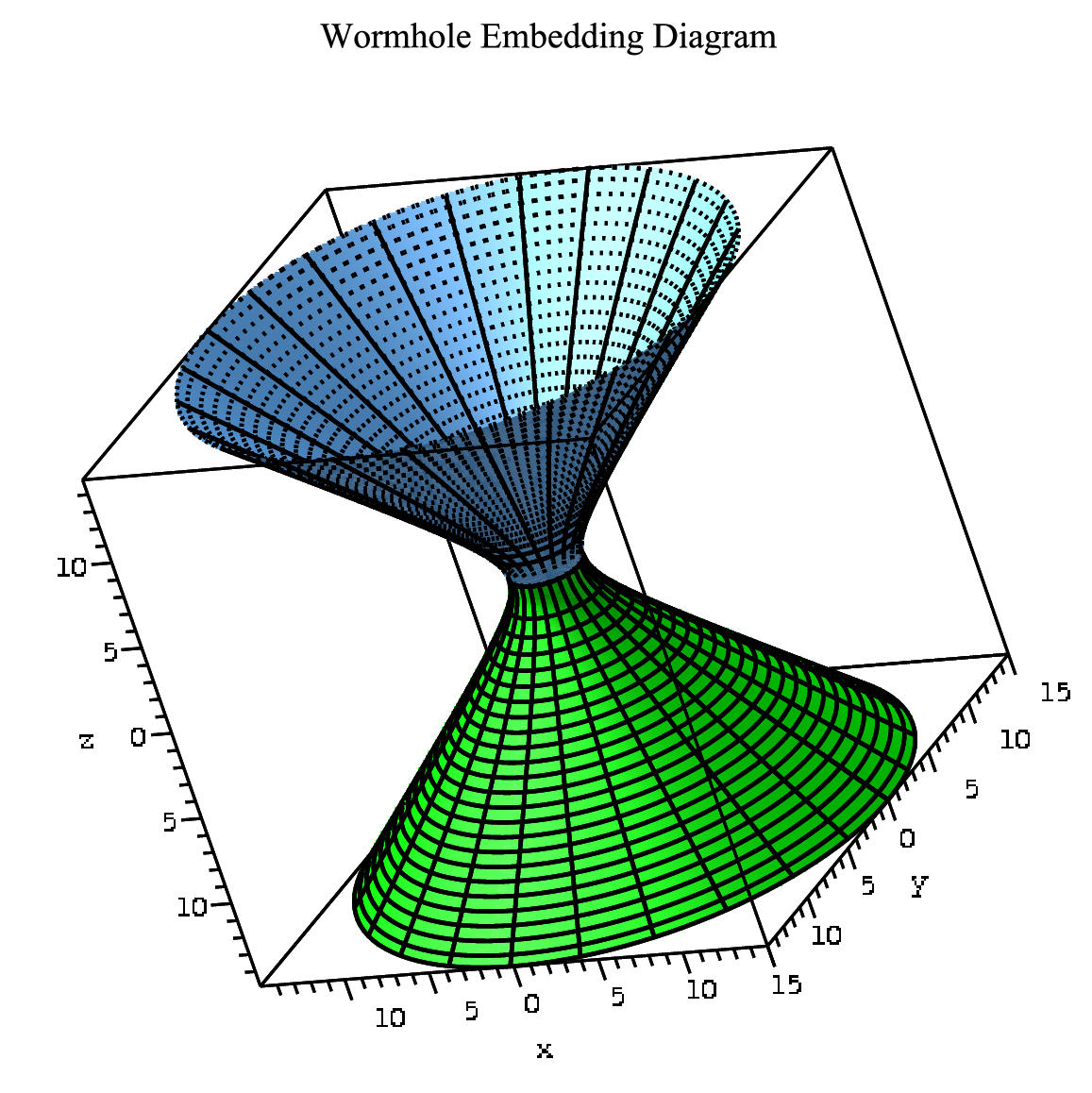}}
\caption{
{\it{
Embedding structure of the rotating traversable wormhole geometry for the 
representative configuration with $m=1$. 
Panel (a) displays the embedding function 
$z(r)=\pm\sqrt{r^2-r_0^2}$, showing the two symmetric branches connected at the 
throat radius $r=r_0$. 
Panel (b) presents the behavior of the derivative $\frac{dz}{dr}$, which 
diverges near the throat, indicating the characteristic vertical slope 
associated with the flare-out condition. 
Panel (c) illustrates the corresponding three-dimensional embedding surface 
obtained by revolving the embedding curve around the symmetry axis, revealing 
the smooth bridge structure connecting two asymptotically flat regions of 
spacetime. 
The figure demonstrates the regular and traversable geometrical character of 
the obtained wormhole solution.}}
}\label{fig:embedding}
\end{figure}

The embedding surface corresponding to the obtained solution is illustrated in 
Fig.~\ref{fig:embedding}. The figure exhibits the characteristic bridge 
structure of the wormhole geometry, with two symmetric asymptotically flat 
regions connected through the throat at $r=r_0$.

\subsection{Preliminary stability analysis}
\label{subsec:stability}

To assess the physical viability of the obtained wormhole solutions, it is 
important to examine their behavior under small perturbations. A common and 
effective approach consists in analyzing the sound-speed propagation within the 
anisotropic matter distribution supporting the wormhole geometry.

For an anisotropic fluid, the radial and tangential sound speeds are defined as
\begin{equation}\label{cas}
v_{sr}^2=\frac{dp_r}{d\rho},
\qquad
v_{st}^2=\frac{dp_t}{d\rho}.
\end{equation}
The causality condition requires that both sound speeds satisfy
\begin{equation}
0\leq v_{sr}^2 \leq 1,
\qquad
0\leq v_{st}^2 \leq 1,
\end{equation}
ensuring that perturbations propagate at subluminal velocities 
\cite{Herrera:1992lwz,Abreu:2007ew}.

For the representative configuration considered in the previous section one 
immediately obtains
$
v_{sr}^2
=
\frac{dp_r}{d\rho}
=
1,
$ and $
v_{st}^2
=
\frac{dp_t}{d\rho}
=
n
=
0.5,
$
thus  both sound speeds lie within the physically acceptable interval,
showing that the obtained configuration satisfies the standard causality 
requirements.

Another useful criterion for anisotropic systems is the Herrera cracking 
condition \cite{Herrera:1992lwz}, according to which potentially stable 
configurations satisfy
\begin{equation}
|v_{st}^2-v_{sr}^2|\leq 1.
\end{equation}
For the present model we find 
$
|v_{st}^2-v_{sr}^2|
=
|0.5-1|
=
0.5<1,
$
which indicates the absence of cracking instabilities.
In addition, the positivity and monotonic decrease of the energy density and 
pressure components contribute to the overall physical consistency of the 
configuration. The absence of divergences for $r>r_0$ ensures that the matter 
distribution remains regular throughout the spacetime exterior to the throat.

We mention   that the stability properties obtained here should be 
viewed as preliminary local indicators of physical viability. A complete 
stability analysis would require the study of dynamical perturbations of both 
the geometry and matter sector beyond the present effective approach. 
Nevertheless, the combined behavior of the sound-speed conditions and cracking 
criterion suggests that the obtained rotating wormhole solutions constitute 
physically consistent configurations within the considered $f(R,T)$ modified 
gravity framework.

\section{Particle dynamics and rotational effects}
\label{sec:dynamics}

We now investigate the dynamical properties of the obtained rotating wormhole 
configurations and the effect of rotation on particle motion. We give 
particular 
attention to frame-dragging effects, effective potentials, and the 
modifications induced by the matter-geometry coupling. The analysis of timelike 
and null trajectories further 
provides important insight into the physical and observational behavior of the 
wormhole spacetime.

\subsection{Effective geodesic motion}
\label{subsec:geodesics}

The motion of test particles provides important information regarding the 
physical and observational properties of rotating compact spacetimes. In the 
framework of $f(R,T)$ gravity, the matter-geometry coupling generally induces 
non-geodesic corrections through the non-conservation of the energy-momentum 
tensor. Nevertheless, before incorporating these effects explicitly, it is 
useful to first investigate the leading geometrical contributions within an 
effective geodesic approximation 
\cite{Liang:2025hzr,AraujoFilho:2025hnf,Hazarika:2024cji}. This approach 
captures the dominant effects of the rotating wormhole geometry,   while 
remaining 
analytically tractable.

Within this approximation, test particles satisfy the geodesic equation
\begin{equation}
\frac{d^2 x^\mu}{d\tau^2}
+
\Gamma^\mu_{\;\alpha\beta}
\frac{dx^\alpha}{d\tau}
\frac{dx^\beta}{d\tau}
=
0,
\end{equation}
where $\tau$ is the affine parameter. Restricting the analysis to the 
equatorial plane $(\theta=\pi/2)$ of the slowly rotating metric 
\eqref{slow_metric}, stationarity and axisymmetry imply the existence of two 
conserved quantities associated with the Killing vectors $\partial_t$ and 
$\partial_\phi$, namely the energy $E$ and angular momentum $L$:
\begin{align}
E &=
-g_{tt}\dot{t}
-g_{t\phi}\dot{\phi},
\\
L &=
g_{\phi\phi}\dot{\phi}
+g_{t\phi}\dot{t}.
\end{align}

Using the metric components, one obtains
\begin{align}
E &=
N^2(r)\dot{t}
+r^2\omega(r)\dot{\phi},
\\
L &=
r^2\dot{\phi}
-r^2\omega(r)\dot{t}.
\end{align}
Therefore, to first order in the slow-rotation approximation, we find
\begin{align}\label{phi}
\dot{t}
&=
\frac{E-\omega(r)L}{N^2(r)},
\\
\dot{\phi}
&=
\frac{L}{r^2}
+
\frac{\omega(r)E}{N^2(r)}.
\label{phi22}
\end{align}

Now, the radial motion follows from the normalization condition
\begin{equation}
g_{\mu\nu}\dot{x}^\mu\dot{x}^\nu
=
-\kappa,
\qquad
\kappa=
\begin{cases}
1, & \text{timelike trajectories},
\\
0, & \text{null trajectories}.
\end{cases}
\end{equation}
Substituting Eqs.~\eqref{phi} and \eqref{phi22} into the metric relation yields
\begin{equation}
\frac{\dot{r}^{\,2}}{1-\frac{b(r)}{r}}
=
\frac{E(E-2\omega(r)L)}{N^2(r)}
-
\frac{L^2}{r^2}
-
\kappa .
\label{ELRauxrel}
\end{equation}
This equation can be written in the form
\begin{equation}\label{pot1}
\dot{r}^2+V_{\mathrm{eff}}(r)=0,
\end{equation}
where we have defined the effective potential as
\begin{equation}\label{pot2}
V_{\mathrm{eff}}(r)
=
-\left(1-\frac{b(r)}{r}\right)
\left[
\frac{E(E-2\omega(r)L)}{N^2(r)}
-\frac{L^2}{r^2}
-\kappa
\right].
\end{equation}

Relation \eqref{pot2} embeds the combined effects of the wormhole 
geometry and rotational frame dragging on particle motion. In particular, the 
coupling term proportional to $\omega(r)EL$ modifies the orbital structure 
relative to the static case. The effective potential remains finite at the 
throat, allowing particles to traverse the wormhole provided that 
$\dot{r}^2\ge0$ at $r=r_0$.

\begin{figure}[H]
\centering
\subfigure[Effective potential $V_{\mathrm{eff}}(r)$]{
\label{fig:pot}
\includegraphics[width=0.27\linewidth]
{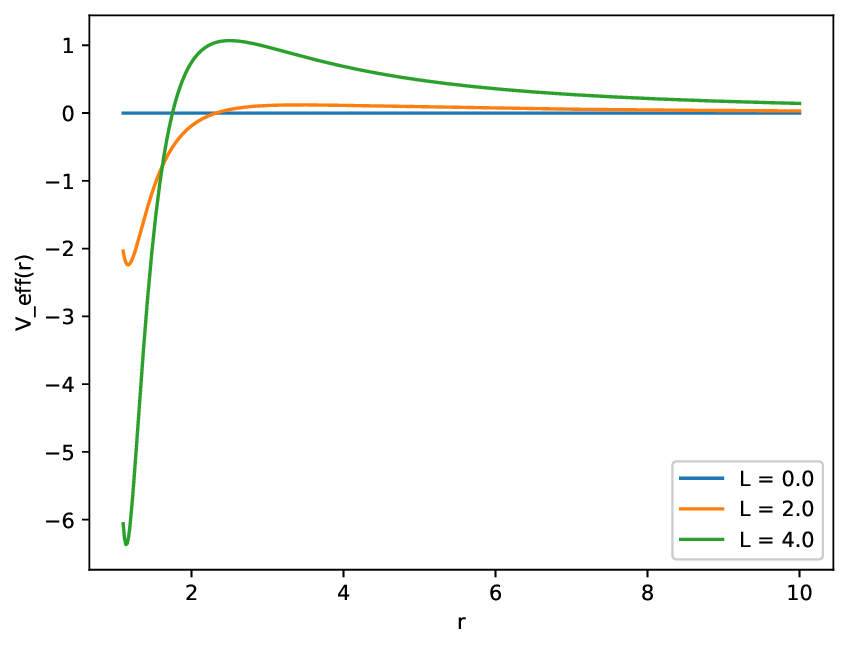}}
\hfill
\subfigure[Orbital trajectory]{
\label{fig:orb}
\includegraphics[width=0.27\linewidth]{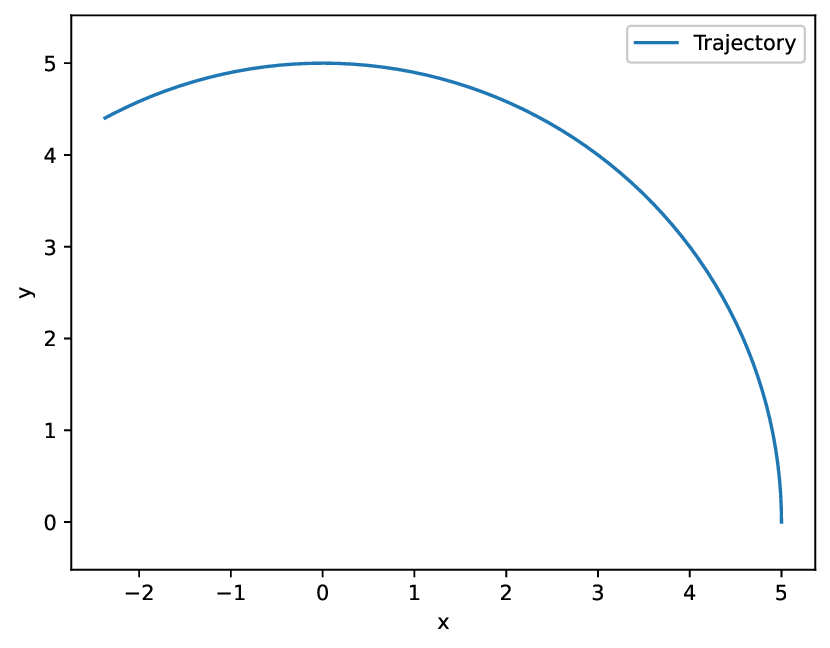}
}
\hfill
\subfigure[Radial velocity profile $\dot{r}$]{
\label{fig:rad}
\includegraphics[width=0.27\linewidth]{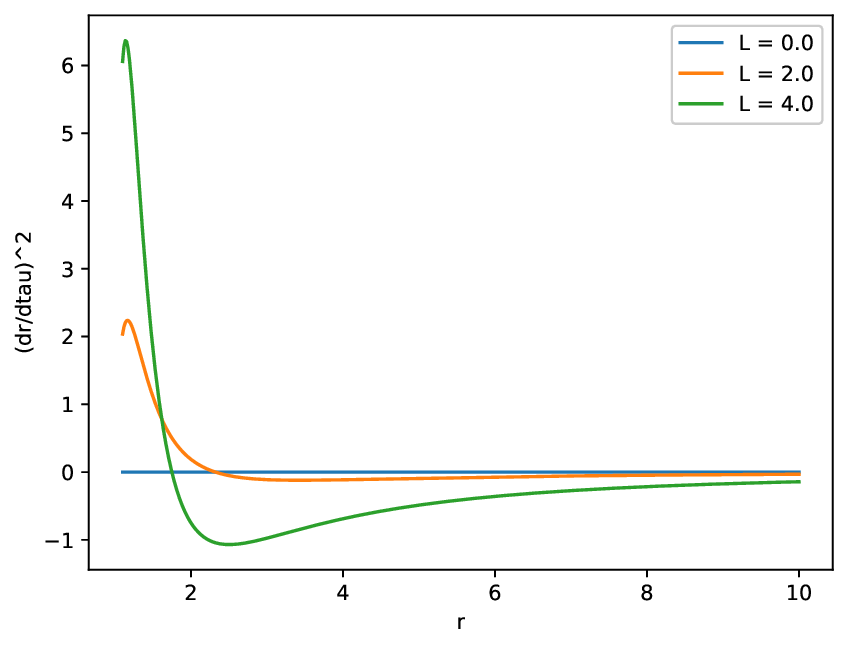}
}\caption{
{\it{
Geodesic motion and effective particle dynamics in the rotating wormhole 
spacetime. 
Panel (a) presents the effective potential $V_{\mathrm{eff}}(r)$ for different 
values of the rotational parameter, illustrating the modification of stable and 
unstable orbital regions due to frame dragging. 
Panel (b) shows a representative orbital trajectory of a test particle in the 
equatorial plane, demonstrating the influence of the wormhole geometry on 
particle motion. 
Panel (c) displays the radial velocity profile $\dot{r}$ as a function of the 
radial coordinate for different rotational configurations. 
The figure illustrates how the combined effects of wormhole geometry and 
rotation modify the orbital structure and traversability properties of the 
spacetime.}}
}
\label{fig:4}
\end{figure}

The effective potential curves shown in 
Fig.~\ref{fig:4}\subref{fig:pot} distinguish between stable and unstable 
circular trajectories. Local minima of $V_{\mathrm{eff}}(r)$ correspond to 
stable orbits, while local maxima indicate unstable configurations. The turning 
points are determined by
$
\dot{r}^2=0
 \Longleftrightarrow  
V_{\mathrm{eff}}(r)=0,
$
which separate allowed and forbidden regions of motion.
Hence, depending on the initial conditions, particles may remain in bounded 
orbits, 
escape to infinity, or cross the wormhole throat. Additionally,
in 
Fig.~\ref{fig:4}\subref{fig:orb} and 
Fig.~\ref{fig:4}\subref{fig:rad}  we display
representative 
orbital 
trajectories and radial velocity profiles,
illustrating the 
traversable character of 
the rotating wormhole geometry.

\subsection{Frame dragging and effective potential}
\label{subsec:frame_dragging}

An important consequence of rotation is the appearance of frame dragging, which 
originates from the off-diagonal metric component $g_{t\phi}$. In rotating 
spacetimes, inertial frames are forced to co-rotate with the geometry, leading 
to observable modifications in particle trajectories and orbital structure.

The angular velocity of a zero-angular-momentum observer (ZAMO), defined by 
$L=0$, is
\begin{equation}
\Omega_{\mathrm{ZAMO}}
=
\frac{d\phi}{dt}
=
\omega(r).
\end{equation}
Hence, even particles with vanishing angular momentum acquire rotational motion 
due to the dragging of inertial frames. As we showed in Eq.~(\ref{asymptbehav}),
for large radial distances $ r\rightarrow\infty$, the rotational profile 
behaves 
as
$\omega(r)\sim \frac{C}{r^3}$,
which corresponds to the standard Lense-Thirring-type decay characteristic of 
slowly rotating compact objects. Consequently, frame-dragging effects become 
dominant near the wormhole throat while rapidly decreasing at large distances, 
where standard asymptotically flat behavior is recovered.
Furthermore, the effect of rotation is also visible in the deformation of the 
effective 
potential \eqref{pot2}. The coupling term proportional to $\omega(r)EL$ shifts 
the locations of stable and unstable orbits, modifies turning points, and 
changes the structure of bounded trajectories relative to the static case. As a 
result, the rotational properties of the spacetime leave direct imprints on the 
dynamics of massive particles orbiting the wormhole.

The radial profiles depicted in Fig.~\ref{fig:4} demonstrate that increasing 
rotation modifies both the infall and escape behavior of particles. In 
particular, the asymmetry introduced by frame dragging affects the orbital 
structure close to the throat, while the asymptotic decay of $\omega(r)$ 
ensures 
the recovery of standard particle motion at sufficiently large distances.
The frame-dragging effects become particularly important near the throat 
region, where the rotational profile reaches its maximum magnitude. 
Consequently, co-rotating and counter-rotating trajectories experience 
different 
effective gravitational interactions, leading to asymmetric orbital behavior.

\subsection{Non-geodesic motion and extra-force contribution}
\label{subsec:extra_force}

We now extend the previous analysis by incorporating the effects of the 
matter-geometry coupling characteristic of $f(R,T)$ gravity. As discussed in 
Sec.~\ref{subsec:field_equations}, the covariant divergence of the 
energy-momentum tensor does not vanish in general, implying that the motion of 
massive test particles deviates from geodesic motion.

The corrected equation of motion is given by~\cite{Harko:2011kv}
\begin{equation}
\frac{d^2x^\mu}{ds^2}
+
\Gamma^\mu_{\alpha\beta}u^\alpha u^\beta
=
f^\mu,
\label{equaext}
\end{equation}
where $f^\mu$ represents the extra force induced by the matter-geometry 
coupling.
For a perfect-fluid description, this force takes the form
\begin{equation}\label{ext}
f^\mu=
\frac{8\pi}
{(\rho+p)(8\pi+f_T)}
\left(g^{\mu\nu}-u^\mu u^\nu\right)\nabla_\nu p .
\end{equation}
In the pressureless limit the extra force vanishes, and the motion reduces to 
standard geodesic motion as in GR.

Although the original expression is usually derived for perfect fluids, it can 
be generalized phenomenologically to anisotropic configurations through an 
effective average pressure.
In the present model, the matter source is anisotropic, and thus we 
introduce 
the average pressure
\begin{equation}
p=\frac{p_r+2p_t}{3}.
\end{equation}
Using the relation
 (\ref{eos}),
together with Eq.~(\ref{rho_pr_pt_general}),
we obtain
\begin{equation}
p(r)
=
-\frac{(1+2n)b(r)}
{3(8\pi+\lambda)r^3}.
\end{equation}
Thus, for the shape function (\ref{shape_general})
the pressure gradient and the combination $\rho+p$ respectively become
\begin{equation}
\frac{dp}{dr}
=
\frac{(1+2n)(m+3)b(r)}
{3(8\pi+\lambda)r^4},
\end{equation}
and
\begin{equation}
\rho+p
=
-\frac{(3m+1+2n)b(r)}
{3(8\pi+\lambda)r^3}.
\end{equation}
Finally, substituting into Eq.~\eqref{ext}, the extra force is obtained as
\begin{equation}
f^\mu
=
-\frac{8\pi(1+2n)(m+3)}
{(8\pi+\lambda)(3m+1+2n)r}
\left(g^{\mu r}-u^\mu u^r\right),
\end{equation}
and therefore  the corrected equation of motion  (\ref{equaext}) 
  takes the form
\begin{equation}
\frac{d^2x^\mu}{ds^2}
+
\Gamma^\mu_{\alpha\beta}u^\alpha u^\beta
=
-\frac{8\pi(1+2n)(m+3)}
{(8\pi+\lambda)(3m+1+2n)r}
\left(g^{\mu r}-u^\mu u^r\right).
\end{equation}
Note that for static radial motion with $u^r=0$, the force reduces to the 
purely radial 
component
\begin{equation}
f^r
=
-\frac{8\pi(1+2n)(m+3)}
{(8\pi+\lambda)(3m+1+2n)r}
\left(1-\frac{b(r)}{r}\right).
\end{equation}
Lastly, for the representative configuration
$
m=1,
$ $
n=\frac{1}{2},
$ $
\lambda=-30,
$ considered in the previous section,
we find 
$
f^r
=
\frac{64\pi}
{5(60-8\pi)r}
\left(1-\frac{r_0^2}{r^2}\right)$.

\begin{figure}[H]
\centering
\subfigure[~Radial extra force $f^r(r)$ for $m=1$, $n=0.5$, and 
$\lambda=-30$]{\label{fig:T1}
\includegraphics[width=0.36\textwidth]{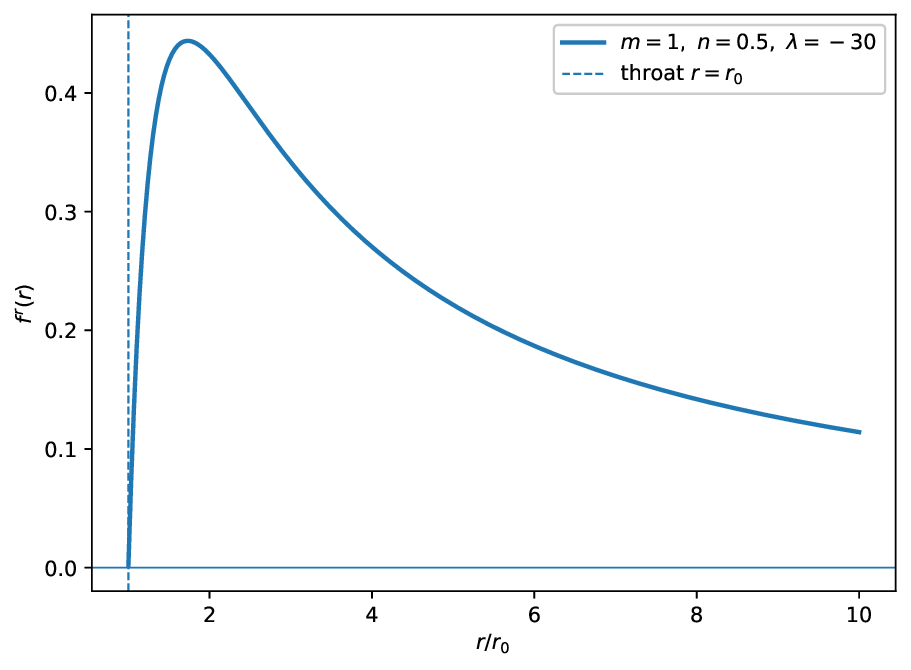}}\hspace{
1cm}
\subfigure[~Effect of the coupling parameter $\lambda$ on the extra 
force]{\label{fig:T2}\includegraphics[width=0.36\textwidth]{
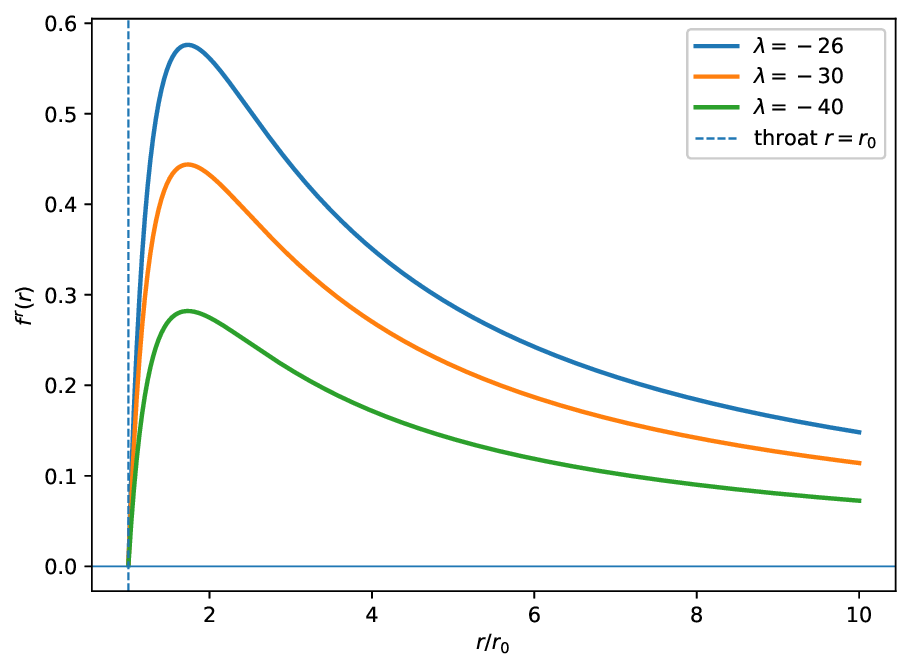}}
\subfigure[~Effect of the anisotropy parameter $n$ on the extra 
force]{\label{fig:T3}\includegraphics[width=0.36\textwidth]{
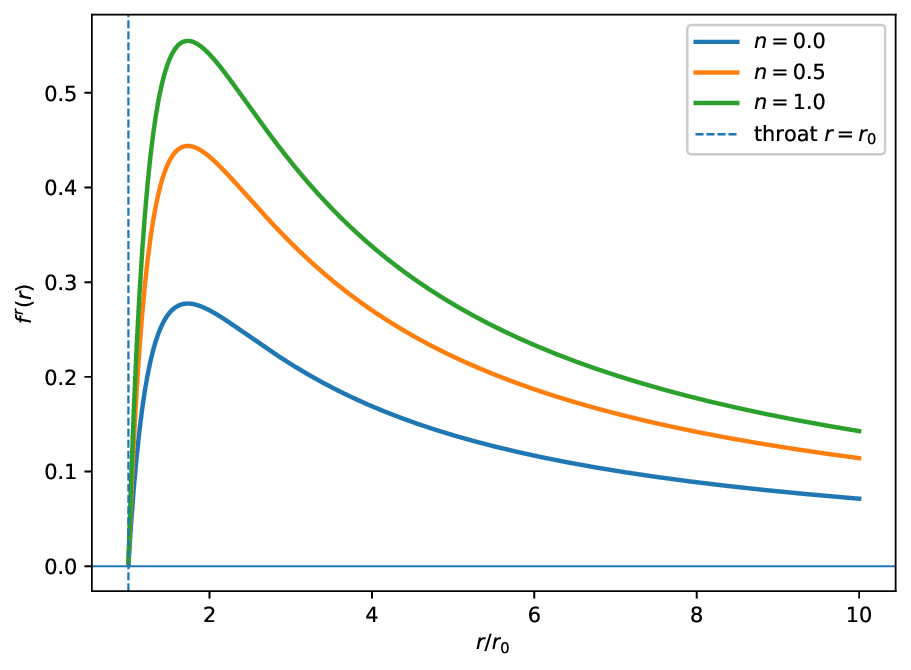}}\hspace{1cm}
\subfigure[~Effective potential with and without the extra 
force]{\label{fig:T4}\includegraphics[width=0.36\textwidth]{
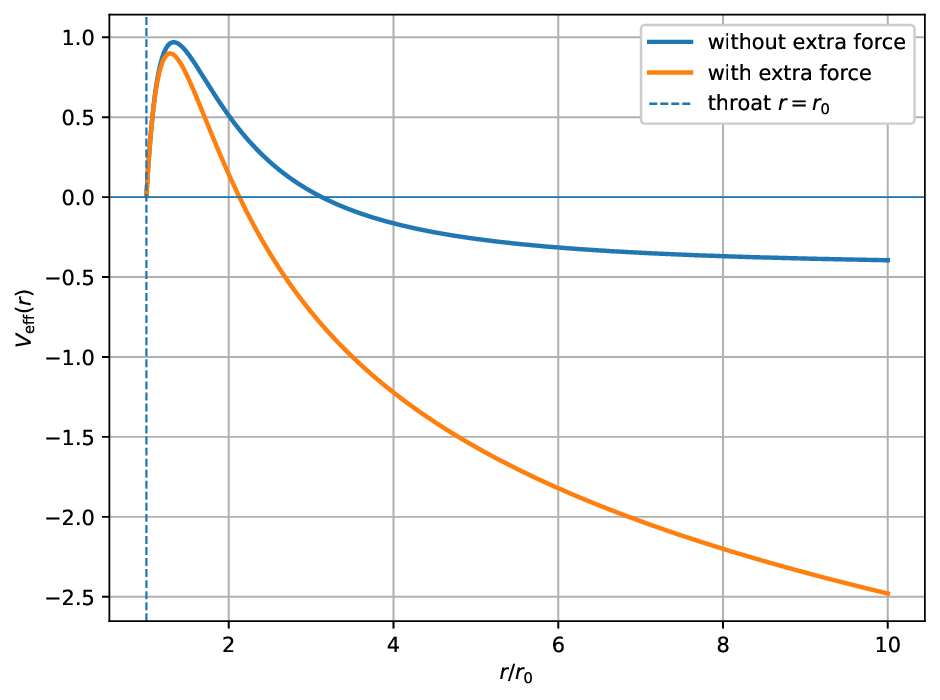}}
\caption{
{\it{
Behavior of the extra-force contribution induced by the matter-geometry 
coupling in $f(R,T)$ gravity and its influence on test-particle dynamics. 
Panel (a) presents the radial extra force $f^{r}(r)$ for the representative 
rotating wormhole configuration with $m=1$, $n=0.5$, and $\lambda=-30$, showing 
that the force vanishes at the throat and decreases at large distances. 
Panel (b) illustrates the dependence of the extra force on the coupling 
parameter $\lambda$, demonstrating the sensitivity of the particle dynamics to 
the strength of the matter-geometry interaction. 
Panel (c) displays the effect of the anisotropy parameter $n$ on the radial 
force profile. 
Panel (d) compares the effective potential with and without the extra-force 
contribution, highlighting the corrections induced by the non-geodesic motion 
of massive particles. 
The figure illustrates how matter-geometry coupling modifies the orbital 
behavior of particles around the rotating wormhole spacetime.}}
}
\label{fig:extra_force}
\end{figure}

In Figs.~\ref{fig:extra_force}\subref{fig:T1}, 
\ref{fig:extra_force}\subref{fig:T2}
and \ref{fig:extra_force}\subref{fig:T3}, 
we present 
the radial behavior of the extra force for different parameter choices.
As we can see, the force vanishes at the throat 
($r=r_0$), reaches a maximum at finite distance from the throat, and decays as 
$1/r$ at large distances.
Additionally, the presence of the extra force modifies the effective motion of 
massive test 
particles relative to the effective geodesic approximation discussed earlier. 
This effect is illustrated in 
Fig.~\ref{fig:extra_force}\subref{fig:T4}, where the effective potentials with 
and without the extra-force contribution are compared. The matter-geometry 
coupling may introduce additional corrections to the orbital structure 
of 
massive particles around the rotating wormhole geometry.

It is important to emphasize that the extra-force contribution affects only the 
motion of massive particles. The null geodesics and the corresponding shadow 
structure remain determined by the spacetime geometry itself. Consequently, the 
shadow deformation discussed in the following section originates from the 
rotational properties of the wormhole spacetime and the associated 
frame-dragging 
effects encoded in $\omega(r)$.

 \section{Shadows and gravitational lensing}
\label{sec:observations}

We finally investigate the observational signatures associated with the 
obtained 
rotating wormhole geometries. In particular, we analyze null geodesics, shadow 
formation, and gravitational lensing in order to examine how rotation and frame 
dragging modify photon trajectories and potentially distinguish these 
configurations from conventional compact objects. These effects provide an 
important connection between the theoretical properties of the solutions and 
their possible astrophysical observability.

\subsection{Null geodesics and shadow structure}
\label{subsec:shadow}

An important observational feature of compact spacetimes is the shadow produced 
by unstable photon trajectories. In rotating wormhole geometries, the shadow 
structure is directly influenced by the rotational properties of the spacetime, 
leading to potentially observable deviations from the static case.

The shadow of the rotating wormhole spacetime can be investigated through the 
analysis of null geodesics ($\kappa=0$). Using the radial equation obtained in 
Sec.~\ref{subsec:geodesics}, namely, Eq.~(\ref{ELRauxrel}) for $\kappa=0$ 
(since 
we examine the null geodesics), and introducing the impact parameter
\begin{equation}
\xi_c=\frac{L}{E},
\end{equation}
then 
the radial equation can be rewritten as
\begin{equation}
\frac{\dot{r}^2}
{E^2\left(1-\frac{b(r)}{r}\right)}
=
\frac{1-2\omega(r)\xi_c}{N^2(r)}
-
\frac{\xi_c^2}{r^2}.
\end{equation}

The boundary of the shadow is determined by unstable circular photon orbits. 
These trajectories satisfy the conditions
\begin{equation}\label{R}
\mathcal{R}(r,\xi_c)=0,
\qquad
\frac{d\mathcal{R}(r,\xi_c)}{dr}=0,
\end{equation}
where
\begin{equation}
\mathcal{R}(r,\xi_c)
=
\frac{1-2\omega(r)\xi_c}{N^2(r)}
-
\frac{\xi_c^2}{r^2}.
\end{equation}

\begin{figure}[H]
\centering
\subfigure[Shadow without extra force]{
\label{fig:vef}
\includegraphics[width=0.43\linewidth]{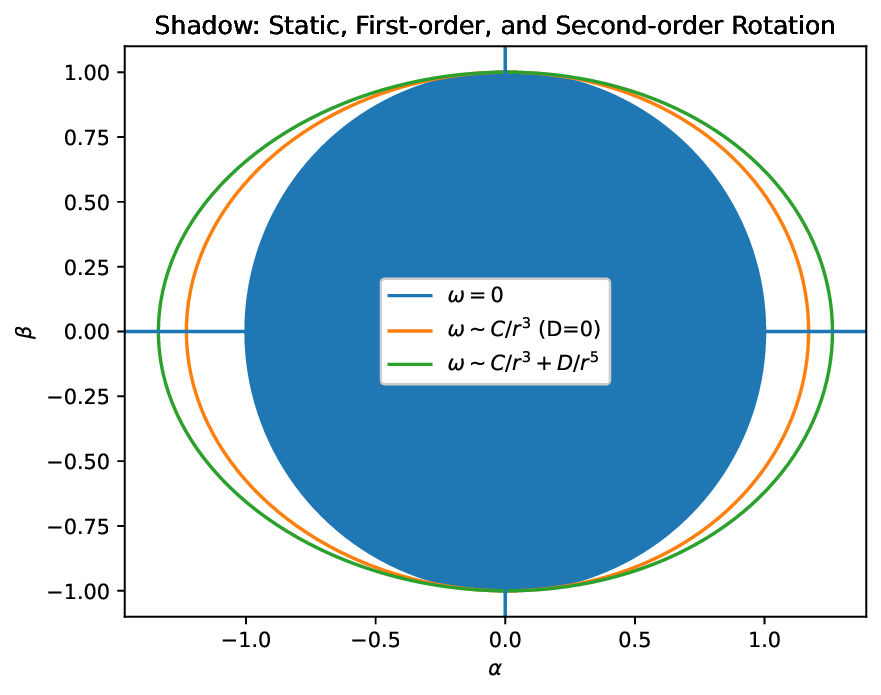}
}
\subfigure[Comparison of shadow deformation for different rotational profiles]{
\label{fig:nvef}
\includegraphics[width=0.43\linewidth]{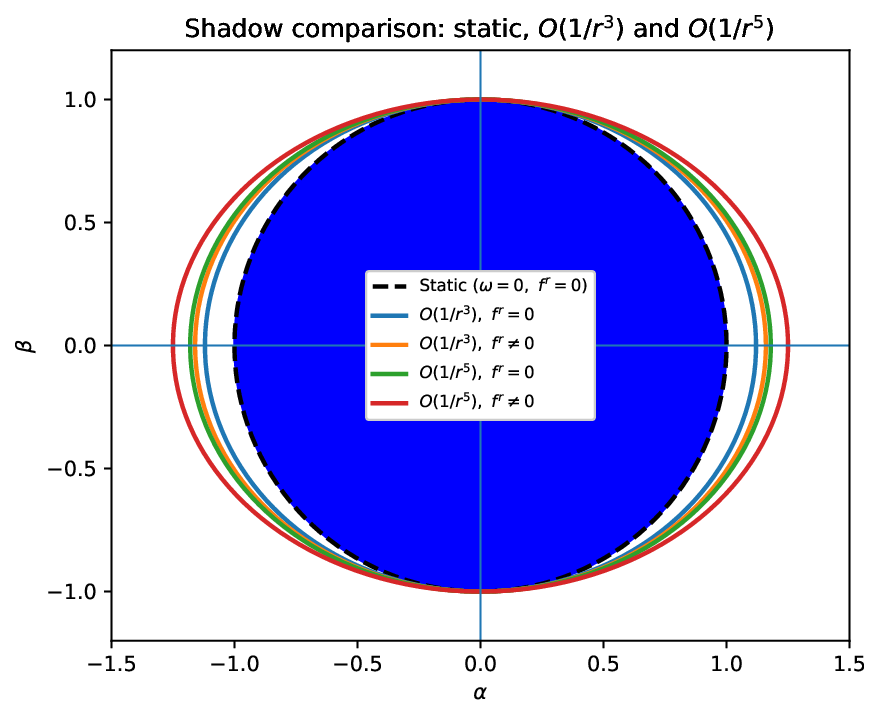}
}
\caption{
{\it{
Shadow structure of the rotating wormhole spacetime in the slow-rotation 
approximation. 
Panel (a) presents the shadow corresponding to the static and rotating 
configurations, illustrating the deformation induced by frame-dragging effects. 
Panel (b) compares the shadow profiles for different rotational contributions, 
showing the asymmetry between co-rotating and counter-rotating photon 
trajectories. 
The static case ($\omega=0$) produces a symmetric circular shadow, while 
rotation generates observable distortions of the shadow boundary through the 
coupling between photon motion and the rotational profile $\omega(r)$. 
The additional force arising from the matter-geometry coupling in $f(R,T)$ 
gravity affects only massive particles and therefore does not modify the shadow 
structure, which is determined exclusively by null geodesics. 
The numerical values used are $C=0.25$, $D=0.08$, and $r_0=1$.}}
}
\label{fig:shadow}
\end{figure}

The above conditions determine the critical values of the impact 
parameter that 
define the apparent boundary of the shadow for a distant observer. In the 
static 
limit $\omega(r)\rightarrow0$, the shadow becomes symmetric. However, rotation 
introduces a coupling between photon angular momentum and the rotational 
profile 
$\omega(r)$, leading to frame-dragging-induced distortions of the shadow shape.
In particular, the rotational contribution modifies the location of unstable 
photon orbits and produces an asymmetry between co-rotating and 
counter-rotating photon trajectories. The resulting deformation of the shadow 
is 
illustrated in Fig.~\ref{fig:shadow}, where the symmetric 
shadow 
of the static configuration is compared with the corresponding slowly rotating 
case.
It is important to note that, although massive test particles are affected 
by the extra-force contribution discussed in 
Sec.~\ref{subsec:extra_force}, the null geodesics remain determined by the 
spacetime 
geometry itself. Consequently, the shadow deformation originates from the 
rotational properties of the wormhole geometry and the associated 
frame-dragging 
effects encoded in $\omega(r)$.

We mention that the present analysis is intended to provide a qualitative 
illustration of 
the leading rotational effects rather than a full observational reconstruction, 
and it is performed within the slow-rotation approximation, where 
higher-order rotational corrections are neglected. Furthermore, the geodesic 
analysis has been restricted to the equatorial plane, which captures the 
leading 
qualitative properties of the shadow while remaining analytically tractable. A 
complete reconstruction of the shadow would require the study of general 
non-equatorial null geodesics beyond the present perturbative treatment, and it 
is left for a future project.

 \subsection{Weak-field lensing signatures}
\label{subsec:lensing}

Gravitational lensing constitutes another important observational probe of the 
wormhole geometry. In particular, the deflection of light rays in the vicinity 
of rotating compact objects can reveal characteristic signatures associated 
with 
frame dragging and the underlying spacetime structure.

Using Eq.~\eqref{R}, the deflection angle of light can be formally written as
\begin{equation}\label{alpha}
\alpha(r_0)
=
2\int_{r_0}^{\infty}
\frac{dr}{\sqrt{\mathcal{R}(r,\xi_c)}}
-\pi,
\end{equation}
where $r_0$ denotes the distance of the closest approach and 
$\mathcal{R}(r,\xi_c)$ is the radial function introduced previously.
The rotational function $\omega(r)$ modifies the bending angle through 
frame-dragging effects, leading to asymmetric light deflection between 
co-rotating and counter-rotating photon trajectories. Additional insight can be 
obtained in the weak-field regime,
$
r_0 \gg r_{\mathrm{th}},
$
where $r_{\mathrm{th}}$ is the throat radius. In this limit,
$
\frac{b(r)}{r}\ll1,
$ $
\omega(r)\ll1,
$
and the radial function becomes
\begin{equation}
\mathcal{R}(r,\xi_c)
\approx
\frac{1}{N^2(r)}
\left(1-2\omega(r)\xi_c\right)
-\frac{\xi_c^2}{r^2}.
\end{equation}
Substituting this expansion into Eq.~\eqref{alpha}, the deflection angle may be 
schematically expressed as
\begin{equation}
\alpha(r_0)
\approx
\alpha_{\mathrm{static}}(r_0)
+
\delta\alpha_{\mathrm{rot}}(r_0),
\end{equation}
where $\alpha_{\mathrm{static}}$ corresponds to the non-rotating wormhole 
contribution, while $\delta\alpha_{\mathrm{rot}}$ denotes the leading 
rotational correction. In the absence of rotation, one recovers
\begin{equation}
\alpha_{\mathrm{static}}(r_0)
=
2\int_{r_0}^{\infty}
\frac{dr}
{\sqrt{
\frac{1}{N^2(r)}
-
\frac{\xi_c^2}{r^2}
}}
-\pi.
\end{equation}

\begin{figure*}[t]
\centering

\begin{minipage}[b]{0.29\textwidth}
    \centering
    \subfigure[Deflection 
angle]{\label{fig:T66}\includegraphics[width=\textwidth,height=4.2cm]{
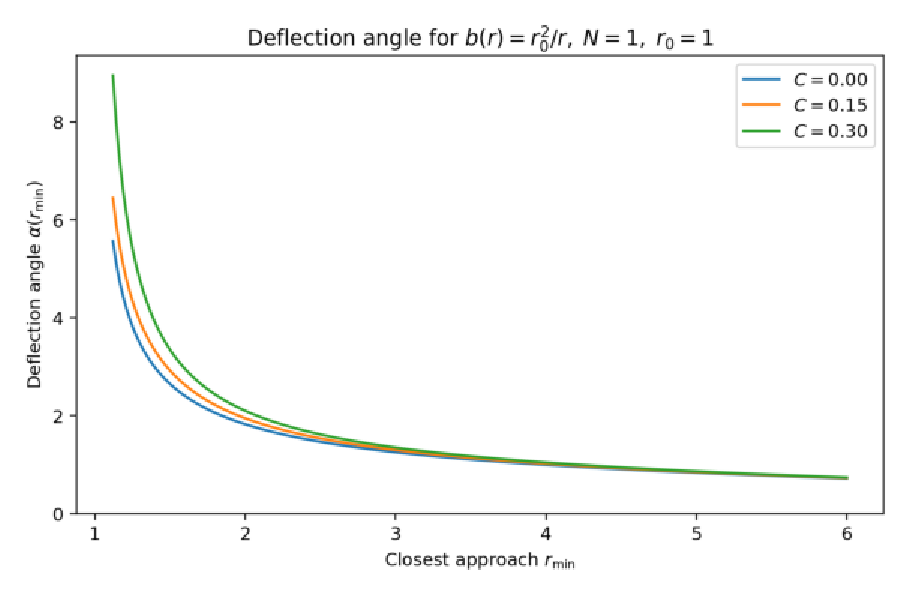}}

    \end{minipage}
\hfill
\begin{minipage}[b]{0.33\textwidth}
    \centering
 \subfigure[Einstein - ring 
deformation]{\label{fig:T77}\includegraphics[width=\textwidth,height=4.4cm]{
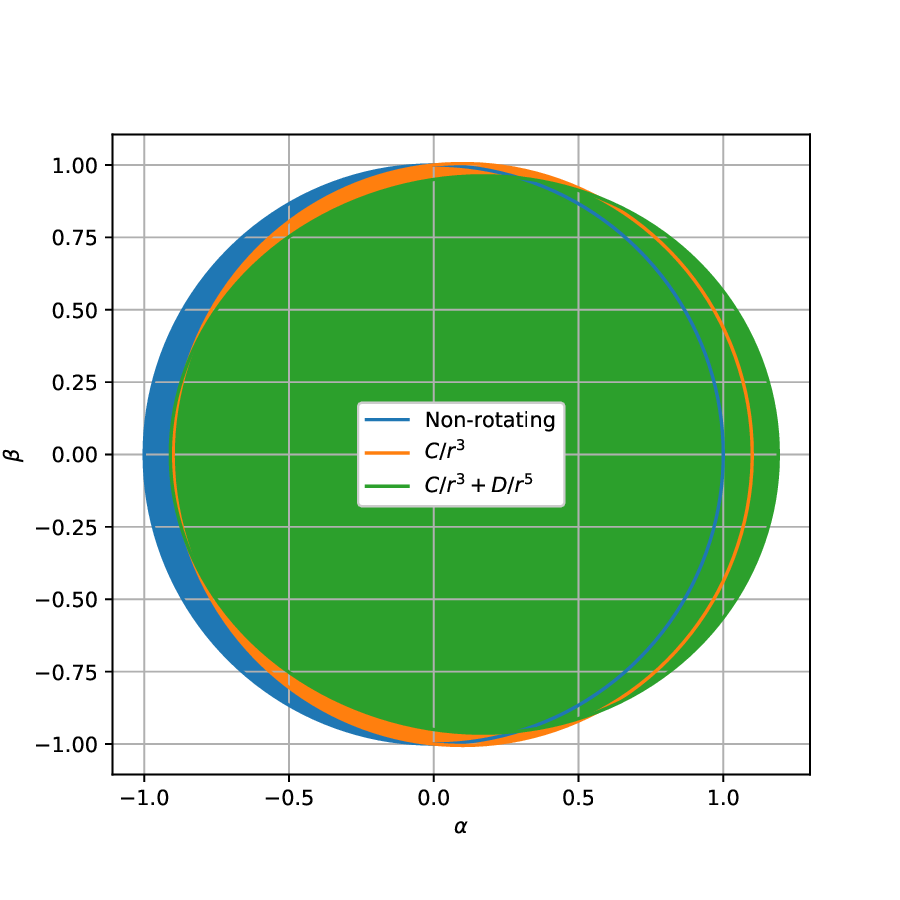}}

\end{minipage}
\hfill
\begin{minipage}[b]{0.31\textwidth}
    \centering
    \subfigure[Radial lensing 
function]{\label{fig:T88}\includegraphics[width=\textwidth,height=4.2cm]{
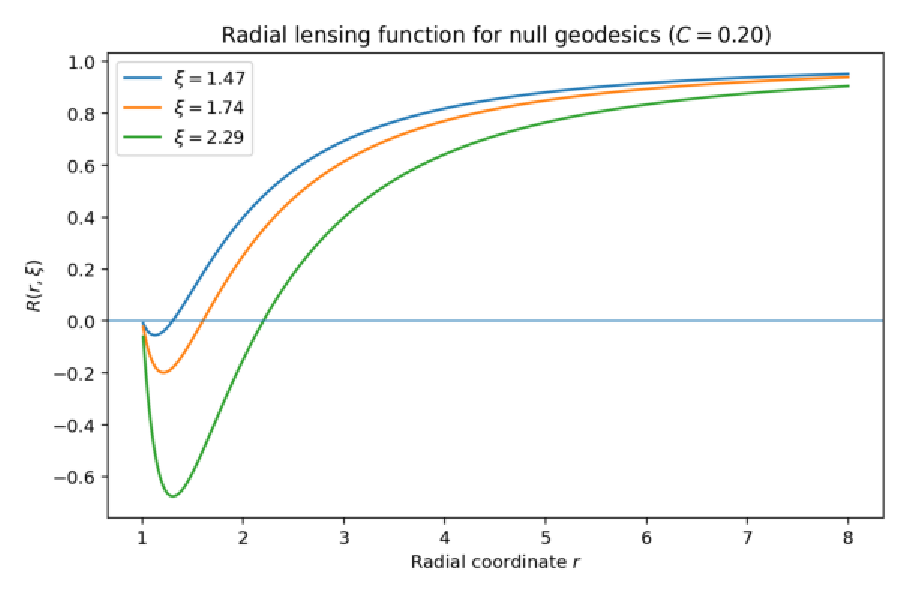}}

\end{minipage}

\caption{
{\it{
Gravitational lensing signatures of the rotating wormhole geometry in the 
slow-rotation approximation. 
Panel (a) presents the deflection angle as a function of the closest-approach 
distance for different values of the rotational parameter $C$, illustrating the 
influence of frame dragging on light bending. 
Panel (b) displays the deformation of the Einstein ring produced by different 
rotational profiles, showing the asymmetry induced by rotation in the lensing 
structure. 
Panel (c) shows the radial lensing function $\mathcal{R}(r,\xi_c)$ governing 
null geodesic propagation in the wormhole spacetime. 
The figure demonstrates how rotation modifies photon trajectories and generates 
characteristic lensing signatures that may distinguish rotating traversable 
wormholes from non-rotating compact configurations.}}
}
\label{fig:lensing}
\end{figure*}

The lensing behavior associated with the obtained rotating wormhole geometry is 
illustrated in Fig.~\ref{fig:lensing}. As shown in the panel (a) of 
Fig.~\ref{fig:lensing}, the deflection 
angle increases as photons propagate closer to the throat, while the rotational 
parameter modifies the bending behavior through frame-dragging effects. 
The panel (b) of Fig.~\ref{fig:lensing} displays the deformation of the Einstein 
ring induced by rotation, 
revealing the asymmetry between co-rotating and counter-rotating photon 
trajectories. Finally, the panel (c) of Fig.~\ref{fig:lensing} presents the 
radial lensing function 
$\mathcal{R}(r,\xi_c)$ governing null geodesic propagation, confirming the 
existence of critical photon trajectories associated with strong-field lensing 
effects.

At large distances, where $N(r)\rightarrow1$ and $b(r)/r\rightarrow0$, the 
spacetime approaches asymptotic flatness and the lensing behavior reduces to 
the standard weak-field regime. The rotational contribution introduces an 
asymmetry in the bending angle through the coupling term 
$\omega(r)\xi_c$, leading to different deflection angles for co-rotating and 
counter-rotating photon trajectories. This effect represents a direct 
manifestation of frame dragging and constitutes a potentially observable 
difference between rotating and non-rotating wormhole configurations.

On the other hand, in the strong-field regime, when photons propagate close to 
the wormhole 
throat, the deflection angle can become arbitrarily large, leading to the 
formation of multiple relativistic images associated with unstable photon 
orbits. The corresponding critical trajectories satisfy Eq.~(\ref{R}). 
Therefore, 
the obtained rotating wormhole geometry  exhibits several 
characteristic lensing signatures, including asymmetric light deflection 
induced 
by frame dragging, modified weak-field bending angles, and shifted 
strong-field relativistic images. These effects may provide a possible 
observational channel for distinguishing rotating traversable wormholes from 
conventional compact objects in future high-resolution astronomical 
observations.

\section{Conclusions}\label{X}
 
\label{sec:conclusions}

Traversable wormholes are among the most intriguing solutions of the 
gravitational field equations, since they provide nontrivial spacetime 
topologies connecting distinct asymptotic regions through a finite throat. 
Beyond their geometrical interest, wormholes play an important role in the 
investigation of strong-field gravity, compact objects, and possible extensions 
of GR. However, within standard GR, traversable wormholes 
generically require exotic matter violating the NEC, 
which significantly limits their physical viability 
\cite{Morris:1988cz,Visser1995}. This difficulty has motivated extensive study 
of wormhole geometries in modified gravity theories, where additional geometric 
contributions may effectively support the wormhole structure without requiring 
pathological matter sources.

In the present work, we have investigated rotating traversable wormholes in the 
framework of $f(R,T)$ gravity, focusing on the linear model 
$f(R,T)=R+\lambda T$. Our motivation was twofold. Firstly, the explicit 
matter-geometry coupling characteristic of $f(R,T)$ gravity provides a natural 
mechanism through which the effective gravitational sector may be modified, 
allowing the existence of wormhole geometries compatible with the standard 
energy conditions. Secondly, rotation constitutes a fundamental property of 
realistic compact astrophysical objects and can significantly influence the 
dynamical and observational properties of wormhole configurations through 
frame-dragging effects, modified orbital structure, and distorted photon 
trajectories.

Working within the slow-rotation approximation, we have constructed rotating 
wormhole configurations supported by an anisotropic matter distribution and 
analyzed their geometrical and physical properties. The obtained solutions are 
asymptotically flat, possess a regular throat satisfying the flare-out 
condition, and remain free of horizons and geometrical singularities. The 
embedding analysis further confirmed the smooth bridge structure connecting two 
asymptotically flat regions. Unlike standard GR wormholes, the matter sector in 
the present framework satisfies both the NEC and the SEC, 
demonstrating that the matter-geometry coupling in $f(R,T)$ gravity can 
effectively support traversable wormhole geometries without requiring exotic 
matter.

An important aspect of the present analysis is the inclusion of rotational 
effects and their impact on particle dynamics. The rotational function 
$\omega(r)$ introduces frame dragging and generates a direct coupling between 
energy and angular momentum in the geodesic equations. Consequently, the 
effective potential and orbital structure differ significantly from the static 
configuration, modifying stable and unstable circular trajectories as well as 
the corresponding turning points. The geodesic analysis further demonstrated 
that particles can traverse the wormhole throat without encountering 
singularities, confirming the traversable character of the obtained solutions. 
In addition, we have examined the non-geodesic corrections induced by the 
matter-geometry coupling, showing that the extra-force contribution may 
further 
modify the effective motion of massive particles 
\cite{Harko:2011kv,AraujoFilho:2025hnf}.

We have also investigated the observational signatures associated with null 
geodesics, gravitational lensing, and shadow formation. The rotational profile 
produces asymmetric photon trajectories through frame dragging, leading to 
deformed shadow structures and modified lensing patterns relative to the static 
case. In the strong-field regime, unstable photon orbits generate large 
deflection angles and relativistic images, while the rotational contribution 
induces asymmetries between co-rotating and counter-rotating photon 
trajectories. Although the present shadow and lensing analyses remain 
qualitative and are restricted to the slow-rotation and equatorial 
approximations, they nevertheless demonstrate that rotating wormholes in 
$f(R,T)$ gravity may exhibit characteristic observational signatures capable of 
distinguishing them from conventional compact objects 
\cite{Bambi:2013nla,Hazarika:2024alm}.

\begin{table*}[t]
\centering
\caption{Comparison between the present rotating wormhole model and 
representative 
wormhole studies in General Relativity and modified gravity theories. 
The Table marks the role of rotation, the behavior of the energy 
conditions, the nature of the matter sources, and the inclusion of dynamical 
and 
observational effects such as geodesic motion, shadow formation, and 
gravitational lensing. 
In contrast to many previous constructions, the present model simultaneously 
incorporates rotation, matter-geometry coupling, non-geodesic effects, and 
observational signatures within a unified $f(R,T)$-gravity framework, while 
supporting traversable wormholes without exotic matter in the usual General 
Relativity sense. 
}\vspace{0.2cm}
\label{tab:comparison_rotating_WH}\resizebox{\columnwidth}{!}{
\begin{tabular}{lcccccc}
\hline\hline
\textbf{Reference} & \textbf{Theory} & \textbf{Rotation} & \textbf{Energy 
Conditions} & \textbf{Matter Type} & \textbf{Geodesics / Shadow} & \textbf{Main 
Features} \\
\hline

Morris \& Thorne (1988)~\cite{Morris:1988cz}
& GR & No
& Violated (NEC)
& Exotic
& No
& First traversable wormhole model \\[3pt]

Teo (1998)~\cite{Teo:1998dp}
& GR & Yes
& Violated (NEC)
& Exotic
& Limited
& Rotating wormhole with frame dragging \\[3pt]

Lobo (2005)~\cite{Lobo:2005vc}
& GR & No
& Violated
& Exotic
& No
& Static wormholes with exotic sources \\[3pt]

Zubair et al. (2016)~\cite{Zubair:2016cde}
& $f(R,T)$
& No
& Can be satisfied
& Anisotropic
& No
& Matter - geometry coupling effects \\[3pt]

Sharif \& Waseem (2018)~\cite{Sharif:2018khl}
& $f(R,T)$
& No
& Partially satisfied
& Anisotropic
& No
& Static anisotropic wormhole solutions \\[3pt]

Hazarika \& Phukon (2025)~\cite{Hazarika:2024cji}
& $f(R,T)$
& No
& Not fully satisfied
& Anisotropic
& Yes
& Shadow analysis in modified gravity \\[3pt]

Araújo Filho et al. (2025)~\cite{AraujoFilho:2025hnf}
& $f(R,T)$
& No
& Model dependent
& Anisotropic
& Yes
& Observational signatures and lensing \\[3pt]
\hline
 
\textbf{Present Work}
& $f(R,T)=$
& Yes
& \textbf{Satisfied}
& Anisotropic
& \textbf{Yes}
& Rotating traversable wormhole without exotic matter; \\

& $R+\lambda T$
& (slow rotation)
& \textbf{(NEC \& SEC)}
& 
& 
& includes stability, geodesics, shadow, and lensing \\

\hline\hline
\end{tabular}}
\end{table*}

The obtained results therefore indicate that rotating wormholes in 
$f(R,T)$ gravity constitute physically consistent and phenomenologically rich 
compact configurations. In contrast to many previously proposed wormhole 
models, the present framework combines regular traversable geometries, 
satisfaction of the standard energy conditions, rotational effects, modified 
particle dynamics, and observationally relevant signatures within a unified 
matter-geometry-coupled gravitational theory.
In order to provide this information in a more transparent way, 
 in 
Table \ref{tab:comparison_rotating_WH}   we 
display the 
comparison. From Table~\ref{tab:comparison_rotating_WH}, one can clearly see the 
 above  distinctions 
relative to both classical GR wormholes and previously studied modified-gravity 
configurations.

Several directions remain open for future investigation. A natural extension of 
the present work would be the study of higher-order rotational corrections 
beyond the slow-rotation approximation, as well as the investigation of general 
non-equatorial trajectories and full numerical ray tracing for shadow and 
lensing reconstruction. It would also be interesting to analyze quasinormal 
modes, accretion processes, and possible gravitational-wave signatures 
associated with rotating wormholes in matter-geometry-coupled gravity. 
These interesting investigations are left for future works.

\begin{acknowledgments}
The work of Kazuharu Bamba was supported in part by the JSPS KAKENHI Grants No. 
24KF0100 and No. 25KF0176.  ENS acknowledges the contribution of the LISA   
CosWG, and of   COST   
Actions 
 CA18108  ``Quantum Gravity Phenomenology in the multi-messenger   approach'',  
CA21136 ``Addressing observational tensions in cosmology with 
  systematics and fundamental physics (CosmoVerse)'', and CA24101
 ``Testing Fundamental Physics with Seismology''.
\end{acknowledgments}

\appendix

\section{Asymptotic behavior of the rotational function and photon circular 
orbits}
\label{app:rotation}

In this appendix, we explain the asymptotic behavior of the rotational function 
and photon circular orbits. 
The rotational profile satisfies
\begin{equation}
\omega(r)
=
C_1
+
C_2
\int
\frac{dr}
{r^4\sqrt{1-\frac{b(r)}{r}}},
\end{equation}
where $C_1$ and $C_2$ are integration constants.
For large radial distances,  and using
$
b(r)=r_0^{m+1}r^{-m},
$
the integrand is expanded as
\begin{equation}
\frac{1}{r^4\sqrt{1-\frac{b(r)}{r}}}
=
r^{-4}
+\frac12 r_0^{m+1}r^{-(m+5)}
+\mathcal{O}(r^{-(2m+6)}).
\end{equation}
Hence, integrating term by term yields
\begin{equation}
\omega(r)
=
C_1
-\frac{C_2}{3r^3}
-\frac{C_2r_0^{m+1}}
{2(m+4)}r^{-(m+4)}
+\cdots .
\end{equation}
Therefore, the leading asymptotic behavior is
\begin{equation}
\omega(r)\sim \frac{C}{r^3},
\qquad
r\to\infty,
\label{asymptoticprof}
\end{equation}
which corresponds to the standard Lense-Thirring-type decay of slowly rotating 
compact configurations.

We now briefly estimate the photon circular orbit and the corresponding 
critical 
impact parameter within the slow-rotation approximation. For null geodesics, 
the radial equation can be written as
\begin{equation}
R(r,\xi)
=
1
-2\omega(r)\xi
-\frac{\xi^2}{r^2},
\end{equation}
where
$
\xi=\frac{L}{E}
$
is the impact parameter.
The circular photon trajectories satisfy
\begin{equation}
R(r_{\rm ph},\xi_c)=0,
\qquad
\left.
\frac{dR(r,\xi_c)}{dr}
\right|_{r=r_{\rm ph}}
=0.
\end{equation}
Using the asymptotic rotational profile in Eq.~(\ref{asymptoticprof}), 
 we find 
\begin{equation}
\xi_c
=
-\frac{3C}{r_{\rm ph}},
\end{equation}
while the photon circular-orbit radius becomes
\begin{equation}
r_{\rm ph}
=
3^{1/4}\sqrt{|C|}.
\end{equation}

In summary, both the photon circular-orbit scale and the critical impact 
parameter depend explicitly on the rotational parameter $C$. These expressions 
remain approximate and are valid within the equatorial and slow-rotation 
approximations adopted throughout the present work.

\end{document}